\pdfoutput=1
\documentclass[letterpaper,twocolumn,10pt]{article}
\usepackage{usenix}
\usepackage{booktabs} % For formal tables
\usepackage{multirow}
\usepackage{makecell}
\usepackage{listings}
\usepackage{bigstrut}
\usepackage{subcaption}
\usepackage{amsmath}
\usepackage{amsfonts}
\usepackage{solidity-highlighting}
\usepackage{xspace}
\usepackage{numprint}
\usepackage{graphicx}
\usepackage{afterpage}
\usepackage{float}
\usepackage{stringstrings}
\usepackage[bottom]{footmisc}
\npthousandsep{,}

\usepackage[T1]{fontenc}
\usepackage{color}
\definecolor{red}{rgb}{1,0,0}
\definecolor{green}{rgb}{0,1,0}
\definecolor{blue}{rgb}{0,0,1}
\definecolor{cyan}{rgb}{0.4,1,1}
\definecolor{orange}{rgb}{1,0.6,0}
\definecolor{dkgreen}{rgb}{0,0.6,0}
\definecolor{dkred}{rgb}{0.6,0,0}
\definecolor{gray}{rgb}{0.5,0.5,0.5}
\definecolor{purple}{rgb}{0.58,0,0.82}

\lstdefinelanguage{json}{
    string=[s]{"}{"},
    stringstyle=\color{black},
    comment=[l]{:},
    commentstyle=\color{black},
}
\lstdefinelanguage{esm}{
    comment=[l]{;},
    commentstyle=\color{gray},
    string=[s]{(}{)},
    stringstyle=\color{gray}
}

\usepackage[font=small,labelfont=bf]{caption}

\newif\ifDRAFT
\DRAFTtrue
\DRAFTfalse

\ifDRAFT
  \newcommand{\fix}[1]{\textcolor{red}{#1}}
  \newcommand{\todo}[1]{{\color{red}\bf\em TODO:\/\@#1}}
  
  \newcommand{\smw}[1]{\todo{Comment by Sam: #1}}
  \newcommand{\empirical}[1]{\textcolor{blue}{#1}}
\else
  \newcommand{\fix}[1]{{}}
  \newcommand{\todo}[1]{{}}
  
  \newcommand{\smw}[1]{{}}
  \newcommand{\empirical}[1]{#1}
\fi

\newcommand{\correctness}{\point{Analysis correctness}\xspace}

\newcommand{\point}[1]{\par\smallskip\noindent\textbf{#1.}}
\newcommand{\addr}[2][\small]{{#1\href{https://etherscan.io/address/#2}{\texttt{#2}}}}
\newcommand{\tx}[2][\small]{{#1\href{https://etherscan.io/tx/#2}{\texttt{\substring{#2}{1}{10}}}}}
\newcommand{\block}[1]{\href{https://etherscan.io/block/#1}{\numprint{#1}}}

\newcommand{\vuln}[1]{\textsf{#1}}
\newcommand{\vre}{\vuln{RE}}
\newcommand{\reentrancy}{Re-Entrancy}
\newcommand{\vue}{\vuln{UE}}
\newcommand{\unhandledexceptions}{Unhandled Exceptions}
\newcommand{\vle}{\vuln{LE}}
\newcommand{\lockedether}{Locked Ether}
\newcommand{\vto}{\vuln{TO}}
\newcommand{\transactionorder}{Transaction Order Dependency}
\newcommand{\vio}{\vuln{IO}}
\newcommand{\integeroverflow}{Integer Overflow}
\newcommand{\vua}{\vuln{UA}}
\newcommand{\unrestrictedaction}{Unrestricted Action}

\newcommand{\op}[1]{\lstinline{#1}}
\newcommand{\dterm}[2]{\lstinline$#1${$\left(#2\right)$}}
\newcommand{\drulesep}{~\lstinline$:-$~}
\newcommand{\dend}{\lstinline$.$}
\newcommand{\dsep}{,~}

\newcommand{\VulnTypes}{six\xspace}
\newcommand{\VulnTypesNum}{6\xspace}
\newcommand{\PapersAnalyzed}{six\xspace}

\newcommand{\ETHRate}{200}
\newcommand{\ToUSD}[1]{\numprint{\the\numexpr #1 * \ETHRate\relax}}

\newcommand{\EtherClaimedVulnerable}{3~million\xspace}
\newcommand{\EtherStake}{3,124,433\xspace}
\newcommand{\EtherClaimedVulnerableUSD}{\ToUSD{3}~million\xspace}

\newcommand{\ExploitedEther}{8,487\xspace}
\newcommand{\ExploitedEtherUSD}{1.7 million\xspace}
\newcommand{\PercentExploitedEther}{0.27\%\xspace}
\newcommand{\PercentExploitedContracts}{1.98\%\xspace}
\newcommand{\VulnerableContracts}{23,327\xspace}
\newcommand{\NumAnalyzedTransactions}{20,241,730\xspace}
\newcommand{\NumExploitedContracts}{463\xspace}

\newenvironment{investigation}[1]{\par\noindent\rule[0.5ex]{\linewidth}{1pt}\par{\noindent{\sf Investigation of the contract at\\ \addr{#1}}}:\\\noindent\rule[0.5ex]{\linewidth}{1pt}}{\\\smallskip{\noindent\rule[0.5ex]{\linewidth}{1pt}}}

\AtEndDocument{\par\leavevmode}

\begin{document}
\sloppy
\title{Smart Contract Vulnerabilities:\\ Vulnerable Does Not Imply Exploited}

\author{
  {\rm Daniel Perez}\\
  Imperial College London
  \and
  {\rm Benjamin Livshits}\\
  Imperial College London
  % copy the following lines to add more authors
  % \and
  % {\rm Name}\\
  % Name Institution
} % end author

\maketitle

\begin{abstract}
  In recent years, we have seen a great deal of both academic and practical interest in the topic of vulnerabilities in smart contracts, particularly those developed for the Ethereum blockchain.
  While most of the work has focused on detecting \emph{vulnerable} contracts, in this paper, we focus on finding how many of these vulnerable contracts have actually been \emph{exploited}.
  We survey the~\VulnerableContracts vulnerable contracts reported by~\PapersAnalyzed recent academic projects and find that, despite the amounts at stake, only~\PercentExploitedContracts of them have been exploited since deployment.
  This corresponds to at most~\ExploitedEther~ETH~(\textasciitilde\ExploitedEtherUSD USD\footnotemark), or only~\PercentExploitedEther of the~\EtherClaimedVulnerable~ETH~(\EtherClaimedVulnerableUSD USD) at stake.
  We explain these results by demonstrating that the funds are very concentrated in a small number of contracts which are \emph{not exploitable} in practice.
\end{abstract}

\footnotetext{We use the exchange rate on 2020-05-16: 1 ETH = \ETHRate{} USD. For consistency, any monetary amounts denominated in USD are based on this rate.}

%\begin{CCSXML}
%<ccs2012>
%<concept>
%<concept_id>10002978.10003029.10011150</concept_id>
%<concept_desc>Security and privacy~Privacy protections</concept_desc>
%<concept_significance>500</concept_significance>
%</concept>
%<concept>
%<concept_id>10002978.10003029.10011703</concept_id>
%<concept_desc>Security and privacy~Usability in security and privacy</concept_desc>
%<concept_significance>300</concept_significance>
%</concept>
%</ccs2012>
%\end{CCSXML}

%\ccsdesc[500]{Security and privacy~Privacy protections}
%\ccsdesc[300]{Security and privacy~Usability in security and privacy}

%\keywords{Smart contracts, Ethereum, vulnerabilities}
\pagestyle{empty}
\section{Introduction}
\label{sec:introduction}

When it comes to vulnerability research, especially as it pertains to software security, it is frequently difficult to estimate what fraction of discovered vulnerabilities are exploited in practice. However, public blockchains, with their immutability, ease of access, and what amounts to a replayable execution log for smart contracts present an excellent opportunity for such an investigation. 
In this work, we aim to contrast the vulnerabilities reported in smart contracts on the Ethereum~\cite{Buterin2014} blockchain with the actual exploitation of these contracts.

We collect the data shared with us by the authors of~\PapersAnalyzed recent papers~\cite{Luu2016a,DBLP:conf/ndss/KalraGDS18,Tsankov2018,Grech2018,Nikolic2018a,217464} focusing on finding smart contract vulnerabilities. These academic datasets are significantly bigger in scale than reports we can find in the wild and because of the sheer number of affected contracts~---~\VulnerableContracts~---~represent an excellent study subject. 

To make our approach more general, we express~\VulnTypes different frequently reported vulnerability classes as Datalog queries computed over relations that represent the state of the Ethereum blockchain. The Datalog-based exploit discovery approach gives more scalability to our process; also, while others have used Datalog for static analysis formulation, we are not aware of it being used to capture the dynamic state of the blockchain over time.

We discover that the amount of smart contract exploitation that occurs in the wild is notably lower than what might be believed, given what is suggested by the sometimes sensational nature of some of the famous cryptocurrency exploits such as TheDAO~\cite{Securities2017} or the Parity wallet~\cite{Breidenbach} bugs.

\point{Contributions} Our contributions are:\itemsep=0pt
\begin{itemize}\itemsep=-2pt
% \item
%     This paper presents the first broadly scoped  analysis of the real-life prominence of security exploits against smart contracts.
\item \textbf{Datalog formulation.}
    We propose a Datalog-based formulation for performing analysis over Ethereum Virtual Machine (EVM) execution traces. We use this highly scalable approach to analyze a total of more than~\empirical{20 million} transactions from the Ethereum blockchain to search for exploits. We highlight that our analyses run \emph{automatically} based on the facts that we extract and the rules defining the vulnerabilities we cover in this paper.

\item \textbf{Experimental evaluation of exploitation.}
    We analyze the vulnerabilities reported in \PapersAnalyzed recently published studies and conclude that, although the number of \emph{vulnerable} contracts and the amount of money at risk is very high, the amount of money actually \emph{exploited} is several orders of magnitude lower.

    We discover that out of~\empirical{\VulnerableContracts} vulnerable contracts worth a total of~\empirical{\EtherStake} ETH,~\empirical{\NumExploitedContracts} contracts may have been exploited for an amount of~\empirical{\ExploitedEther} ETH, which represents only~\empirical{\PercentExploitedEther} of the total amount at stake.

\item \textbf{Proposed explanations.}
    We hypothesize that the main reasons for these vast differences is that the amount of \emph{exploitable} Ether is very low compared to the amount of Ether flagged \emph{vulnerable}.
    Indeed, further analysis of the vulnerable contracts and the Ether they contain suggests that a large majority of Ether is held by only a small number of contracts, and that the vulnerabilities reported on these contracts are either false positives or not exploitable in practice. We also confirm that the set of all contracts on the Ethereum blockchain has a similar distribution of wealth to that in our dataset.
\end{itemize}
To make many of the discussions in this paper more concrete, we present a thorough investigation of the high-value contracts in Appendix~\ref{sec:investigations}. 

%\point{Paper organization}
% The remainder of this paper is organized as follows.
%Sections~\ref{sec:background} and~\ref{sec:related} provide background about
%smart contracts and related work about exploits and tools to prevent them.
% Section~\ref{sec:methodology} presents the methodology used in this paper to
% analyze the real-world impact of the different type of vulnerabilities.

% Sections~\ref{sec:background} and~\ref{sec:related} provides a brief background about tracking and ad-blockers as well as a discussion of the related work.
% Section~\ref{sec:easylistpast} presents a 9-year analysis of EasyList's evolution.
% In Section~\ref{sec:easylistpresent} we present how EasyList rules are applied on websites.
% Section~\ref{sec:applicability} proposes two new blocking strategies to process requests faster.
% Finally, in Section~\ref{sec:threats} we present the limitations,
% and we conclude the paper in Section~\ref{sec:conclusion}.

\section{Background}
\label{sec:background} 
The Ethereum~\cite{Buterin2014} platform allows its users to run ``smart contracts'' on its distributed infrastructure. Ethereum \emph{smart contracts} are programs which define a set of rules for the governing of associated funds, typically written in a Turing-complete programming language called Solidity~\cite{Dannen:2017:IES:3103305}. Solidity is similar to JavaScript, yet some notable differences are that it is strongly-typed and has built-in constructs to interact with the Ethereum platform. Programs written in Solidity are compiled into low-level untyped bytecode to be executed on the Ethereum platform by the Ethereum Virtual Machine (EVM)~\cite{wood2014ethereum}. It is important to note that it is also possible to write EVM contracts without using Solidity.

To execute a smart contract, a sender has to send a transaction to the contract and pay a fee which is derived from the contract's computational cost, measured in units of~\emph{gas}. Each executed instruction consumes an agreed upon amount of gas~\cite{wood2014ethereum}. Consumed gas is credited to the miner of the block containing the transaction, while any unused gas is refunded to the sender. In order to avoid system failure stemming from never-terminating programs, transactions specify a gas limit for contract execution~\cite{DBLP:conf/ndss/0002L20}. An out-of-gas exception is thrown once this limit has been reached.

Smart contracts themselves have the capability to \emph{call} another account present on the Ethereum blockchain. This functionality is overloaded, as it is used both to call a function in another contract and to send Ether (ETH), the underlying currency in Ethereum, to an account. A particularity of how this works in Ethereum is that calls from within a contract, also called \emph{internal transactions}, do not create new transactions and are therefore not directly recorded on-chain. This means that looking at transactions without executing them does not provide enough information to follow the flow of Ether.

\subsection{Smart Contracts Vulnerabilities}
\label{ssec:vulnerability-types}
In this subsection, we briefly review some of the most common vulnerability types that have been researched and reported for EVM-based smart contracts. We provide a two-letter abbreviation for each vulnerability which we shall use throughout the remainder of this paper.

\point{\reentrancy~(\vre)}
When a contract ``calls'' another account, it can choose the amount of gas it allows the called party to use. If the target account is a contract, it will be executed and can use the provided gas budget. If such a contract is malicious and the gas budget is high enough, it can try to call back in the caller --- a re-entrant call. If the caller's implementation is not re-entrant, for example because it did not update his internal state containing balances information, the attacker can use this vulnerability to drain funds out of the vulnerable contract~\cite{Luu2016a,DBLP:conf/ndss/KalraGDS18,Tsankov2018}.
This vulnerability was used in TheDAO exploit~\cite{Securities2017}, essentially causing the Ethereum community to decide to rollback to a previous state using a hard-fork~\cite{mehar2019understanding}. We provide more details about TheDAO exploit in Section~\ref{sec:related}

\point{\unhandledexceptions~(\vue)}
Some low-level operations in Solidity such as \lstinline{send}, which is used to send Ether, do not throw an exception on failure, but rather report the status by returning a boolean. If this return value is unchecked, the caller continues its execution even if the payment failed, which can easily lead to inconsistencies~\cite{DBLP:journals/corr/abs-1809-03981,Luu2016a,Tikhomirov2017,DBLP:conf/ndss/KalraGDS18}.

\point{\lockedether~(\vle)}
Ethereum smart contracts can, as any account on Ethereum, receive Ether. However, there as several reasons for which the received funds might get locked permanently into the contract.

One reason is that the contract may depend on another contract which has been
destructed using the \op{SELFDESTRUCT} instruction of the EVM --- i.e. its code has been removed and its funds transferred. If this was the only way for such a contract to send Ether, it will result in the funds being permanently locked. This is what happened in the Parity Wallet bug in November~2017, locking millions of USD worth of Ether~\cite{Breidenbach}. We provide more details about it in Section~\ref{sec:related}

There are also cases where the contract will \emph{always} run out of gas when trying
to send Ether which could result in locking the contract funds. More details about such issues can be found in~\cite{Grech2018}.

\point{\transactionorder~(\vto)}
In Ethereum, multiple transactions are included in a single block, which means that the state of a contract can be updated multiple times in the same block. If the order of two transactions calling the same smart contract changes the final outcome, an attacker could exploit this property. For example, given a contract which expects participant to submit the solution to a puzzle in exchange for a reward, a malicious contract owner could reduce the amount of the reward when the transaction is submitted.

\point{\integeroverflow~(\vio)}
Integer overflow and underflow is a common type of bug in many programming languages but in the context of Ethereum it can have very severe consequences. For example, if a loop counter were to overflow, creating an infinite loop, the funds of a contract could become completely frozen. This can be exploited by an attacker if he has a way of incrementing the number of iterations of the loop, for example, by registering enough users to trigger an overflow.

\point{\unrestrictedaction~(\vua)}
Contracts often perform authorization, by checking the sender of the message, to restrict the type of action that a user can take.
Typically, only the owner of a contract should be allowed to destroy the contract or set a new owner.
Such an issue can happen not only if the developer forgets to perform critical checks but also if an attacker can execute arbitrary code, for example by being able to control the address of a delegated call~\cite{217464}.

% Typically, only the owner 
% This owner is usually set in the contract constructor but some contracts were found vulnerable because the owner was not initialized correctly, allowing, for example, an attacker to take ownership of the contract.
% A reason for such a bug could be a misnamed function in older versions of Solidity~\cite{Brent2018,Krupp2018}.
% This issue was the root cause of the the Parity wallet bug~\cite{Tsankov2018,Nikolic2018a} which froze more than 500k Ether.

\begin{figure}[tb]
  \footnotesize
  \centering
  \setlength{\tabcolsep}{4.7pt}
  \begin{tabular}{lp{3mm}p{3mm}p{3mm}p{3mm}p{3mm}p{3mm}cc}
    \toprule
    \multirow{2}{*}{\bf Name}	& \multicolumn{6}{c}{\bf Vulnerabilities} & \bf Report & \multirow{2}{*}{\bf  Citation}\\
                              & \vre & \vue & \vle & \vto & \vio & \vua & \bf month &\\
    \midrule
    Oyente         & \checkmark & \checkmark &            & \checkmark & \checkmark &            & 2016-10 & \cite{Luu2016a}       \\ \midrule
    ZEUS           & \checkmark & \checkmark & \checkmark & \checkmark & \checkmark &            & 2018-02 & \cite{DBLP:conf/ndss/KalraGDS18}          \\ \midrule
    Maian          &            &            & \checkmark &            &            & \checkmark & 2018-03 & \cite{Nikolic2018a}   \\ \midrule
    SmartCheck     & \checkmark & \checkmark & \checkmark &            & \checkmark &            & 2018-05 & \cite{Tikhomirov2017} \\ \midrule
    Securify       & \checkmark & \checkmark & \checkmark & \checkmark &            & \checkmark & 2018-06 & \cite{Tsankov2018}    \\ \midrule
    ContractFuzzer & \checkmark & \checkmark &            &            &            &            & 2018-09 & \cite{Jiang2018}      \\ \midrule
    teEther        &            &            &            &            &            & \checkmark & 2018-08 & \cite{217464}      \\ \midrule
    Vandal         & \checkmark & \checkmark &            &            &            &            & 2018-09 & \cite{DBLP:journals/corr/abs-1809-03981}      \\ \midrule
    MadMax         &            &            & \checkmark &            & \checkmark &            & 2018-10 & \cite{Grech2018}      \\
    \bottomrule
  \end{tabular}
  \vskip 1mm
  \caption{A summary of smart contract analysis tools presented in prior work.}
\label{fig:prior-results}
\end{figure}

\subsection{Analysis Tools}
\label{ssec:analysis-tools}
Smart contracts are generally designed to manipulate and \emph{hold} funds denominated in Ether. This makes them very tempting attack targets, as a successful attack may allow the attacker to directly steal funds from the contract. Given the many common vulnerabilities in smart contracts, some of which we described in the previous section, a large number of tools have been developed to find them automatically~\cite{Luu2016a,Tsankov2018,mythril}. Most of these tools analyze either the contract source code or its compiled EVM bytecode and look for known security issues, such as re-entrancy or transaction order dependency vulnerabilities. We present a summary of these different works in Figure~\ref{fig:prior-results}. The second and third columns respectively present the reported number of contracts analyzed and contracts flagged vulnerable in each paper. The ``vulnerabilities'' columns show the type of vulnerabilities that each tool can check for. We present these vulnerabilities in Section~\ref{ssec:vulnerability-types} and give a more detailed description of these tools in Section~\ref{ssec:related-analysis-tools}.

\subsection{Definitions}
\label{ssec:definitions}
We give the definitions used in this paper for the terms \emph{vulnerable}, \emph{exploitable} and \emph{exploited}.
\begin{description}
\item[vulnerable:]
  A contract is vulnerable if it has been flagged by a static analysis tool as such.
  As we will see later, this means that some contracts may be \emph{vulnerable} because of a false-positive.
\item[exploitable:]
  A contract is exploitable if it is vulnerable and the vulnerability could be exploited by an external attacker.
  For example, if the ``vulnerability'' flagged by a tool is in a function which requires to own the contract, it would be \emph{vulnerable} but not \emph{exploitable}.
\item[exploited:]
  A contract is exploited if it received a transaction on Ethereum's main network which triggered one of its vulnerabilities.
  Therefore, a contract can be \emph{vulnerable} or even \emph{exploitable} without having been \emph{exploited}.
\end{description}

\begin{figure}
  \centering
  \setlength{\tabcolsep}{5pt}
  \begin{tabular}{lrrr}
    \toprule
    \bf \multirow{2}{*}{Name} & \bf Contracts & \bf Vulnerabilities & \bf Ether at stake\\
                              & \bf analyzed & \bf found & {\small at time of report} \\
    \midrule
    Oyente & 19,366 & 7,527 & 1,287,032\\
    Zeus & 1,120 & 861 & 671,188\\
    Maian & NA & 2,691 & 15.59 \\
    Securify & 29,694 & 9,185 & 724,306\\
    MadMax & 91,800 & 6,039 & 1,114,958\\
    teEther & 784,344 & 1,532 & 1.55\\
    \bottomrule
  \end{tabular}
  \vskip 1mm
  \caption{Summary of the contracts in our dataset.}
  \label{fig:dataset-stats}
\end{figure}

\section{Dataset}
\label{sec:datasets}
In this paper, we analyze the vulnerable contracts reported by the following~\PapersAnalyzed academic papers:~\cite{Luu2016a},~\cite{DBLP:conf/ndss/KalraGDS18},~\cite{Nikolic2018a},~\cite{Tsankov2018},~\cite{Grech2018} and~\cite{217464}. To collect information about the addresses analyzed and the vulnerabilities found, we reached out to the authors of the different papers.

Oyente~\cite{Luu2016a} data was publicly available~\cite{oyente-benchmarks}. The authors of the other papers were kind enough to provide us with their dataset. We received all the replies within less than a week of contacting the authors.

We also reached out to the authors of~\cite{Tikhomirov2017},~\cite{Jiang2018} and~\cite{DBLP:journals/corr/abs-1809-03981} but could not obtain their dataset, which is why we left these papers out of our analysis.

Our dataset is comprised of a total of \empirical{821,219} contracts, of which \VulnerableContracts contracts have been flagged as vulnerable to at least one of the~\VulnTypes vulnerabilities described in Section~\ref{sec:background}. Although we received the data directly from the authors, the numbers of contracts analyzed usually did not match the data reported in the papers, which we show in Figure~\ref{fig:prior-results}. We believe the two main results for this are: authors improving their tools after the publication and authors not including duplicated contracts in their data they provided us. Therefore, we present the numbers in our dataset, as well as the Ether at stake for vulnerable contracts in Figure~\ref{fig:dataset-stats}. The Ether at stake is computed by summing the balance of all the contracts flagged vulnerable. We use the balance at the time at which each paper was published rather than the current one, as it gives a better sense of the amount of Ether which could potentially have been exploited.

\begin{figure}[tb]
  \begin{subfigure}{0.48\columnwidth}
    \includegraphics[width=\textwidth]{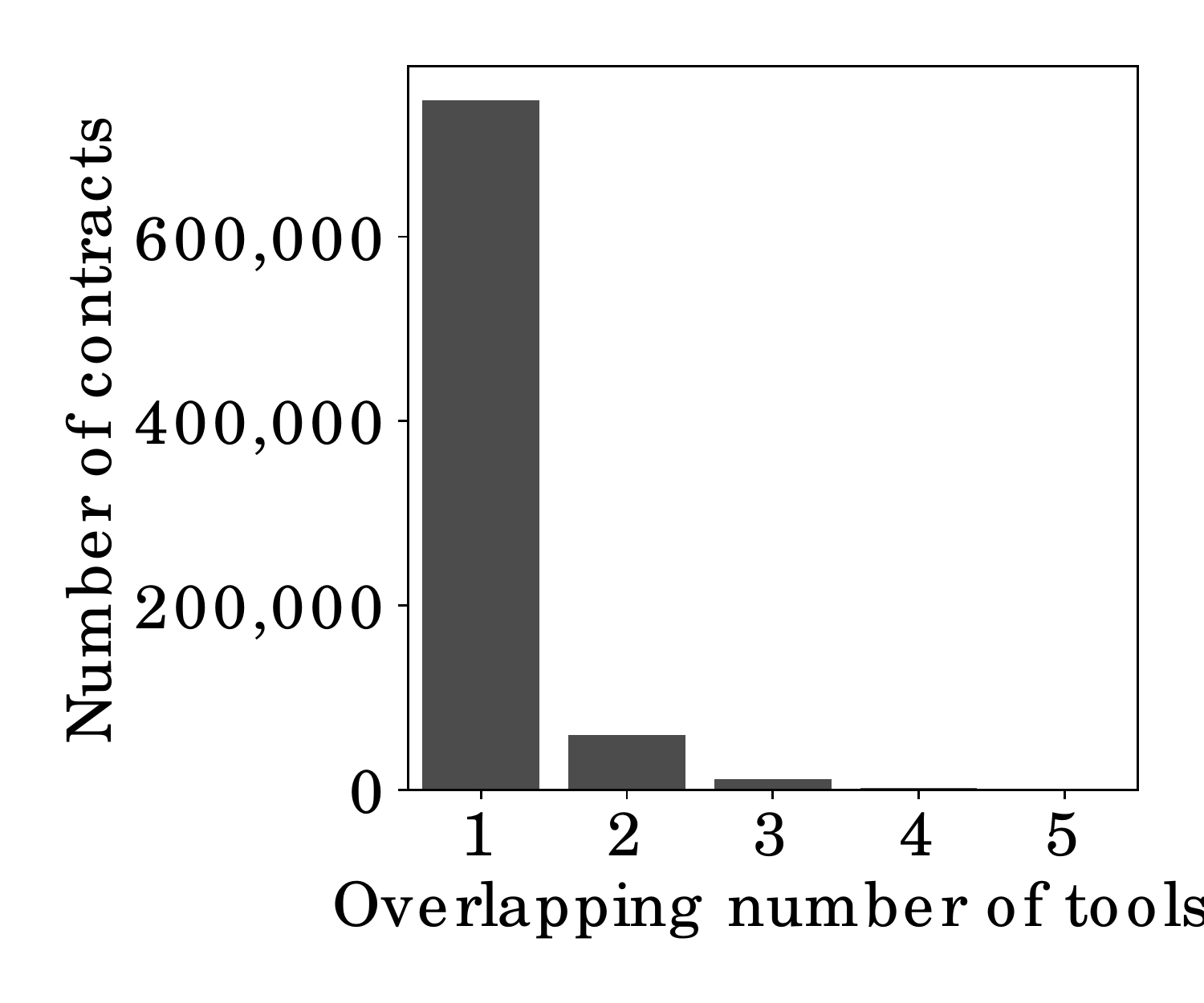}
    \caption{Overlapping contracts\\analyzed.}
    \label{fig:all-overlap}
  \end{subfigure}
  \begin{subfigure}{0.48\columnwidth}
    \includegraphics[width=\textwidth]{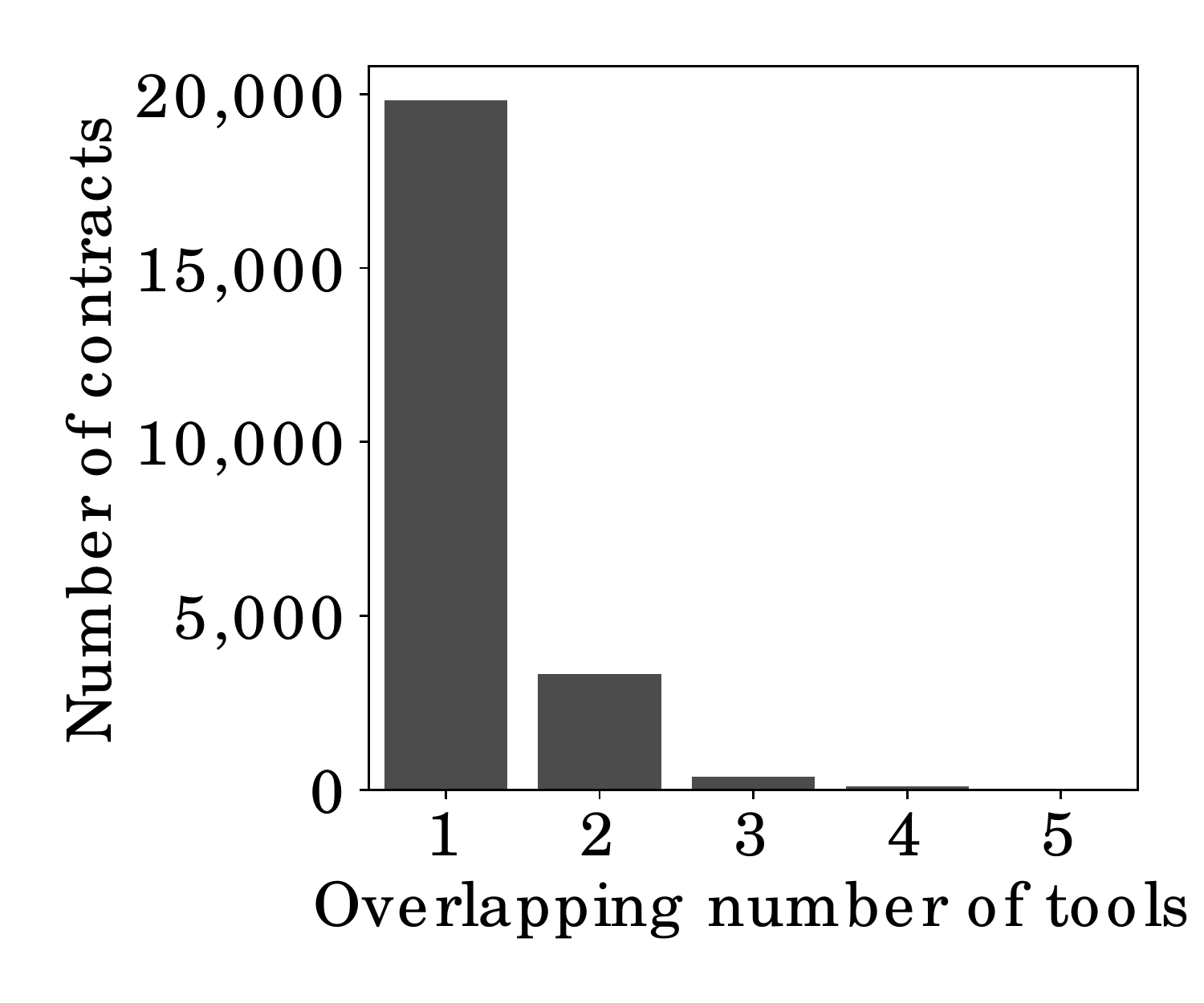}
    \caption{Overlapping vulnerabilities\\\centering flagged.}
    \label{fig:vulnerable-overlap}
  \end{subfigure}
  \caption{Histograms that show the overlap in the contracts analyzed and flagged by examined tools.}
\label{fig:hist-per-tool}
\end{figure}

\begin{figure}[tb]
  \setlength{\tabcolsep}{2pt}
  \centering
  \begin{tabular}{lrrrr}
    \toprule
    \bf Tools & \bf Total & \bf Agreed & \bf Disagreed & \bf \% agreement\\
    \midrule
    Oyente/Securify & 774 & 185 & 589 & 23.9\%\\
    Oyente/Zeus & 104 & 3 & 101 & 2.88\%\\
    Zeus/Securify & 108 & 2 & 106 & 1.85\%\\
    \bottomrule
  \end{tabular}
  \vskip 1mm
  \caption{Agreement among tools for re-entrancy analysis.}
  \label{fig:reentrancy-agreement}
\end{figure}

\begin{figure*}
  \centering
  \setlength{\tabcolsep}{6.5pt}
  \small
  \begin{tabular}{lllllll}
    \toprule
    & \bf Oyente & \bf ZEUS & \bf Securify & \bf MadMax & \bf Maian & \bf teEther\\
    \midrule
    \bf \vre & re-entrancy & re-entrancy & no writes after call & --- & --- & ---\\
    \hline
    \bf \vue & callstack & unchecked send & handled exceptions & --- & --- & ---\\
    \hline
    \bf \vto & concurrency & tx order dependency & transaction ordering dependency & --- & --- & ---\\
    \hline
    \bf \vle & --- & failed send & Ether liquidity & unbounded mass operation & greedy & ---\\
    & & & & wallet griefing\\
    \hline
    \bf \vio & --- & integer overflow & --- & integer overflows & --- & --- \\
    \hline
    \bf \vua & --- & integer overflow & --- & integer overflows & prodigal & exploitable\\
    \bottomrule
  \end{tabular}
  \vskip 1mm
  \caption{Mapping of the different vulnerabilities analyzed.}
  \label{fig:vuln-mapping}
\end{figure*}

\point{Taxonomy}
Rather than reusing existing smart contracts vulnerabilities taxonomies~\cite{10.1007/978-3-662-54455-6_8} as-is, we adapt it to fit the vulnerabilities analyzed by the tools in our dataset. We do not cover vulnerabilities not analyzed by at least \empirical{two} of the \PapersAnalyzed tools. We settle on the~\VulnTypes types of vulnerabilities described in Section~\ref{sec:background}: re-entrancy (\vre), unhandled exception (\vue), locked Ether (\vle), transaction order dependency (\vto), integer overflows (\vio) and unrestricted actions (\vua). As the papers we survey use different terms and slightly different definitions for each of these vulnerabilities, we map the relevant vulnerabilities to one of the \VulnTypes types of vulnerabilities we analyze. We show how we mapped these vulnerabilities in Figure~\ref{fig:vuln-mapping}.

% \point{Excluded data}
% We exclude teEther~\cite{Krupp2018} and Maian~\cite{Nikolic2018a} from our analysis for two reasons. First, the amount of Ether at stake is too low to make an impact on our final results. The amount of Ether at stake for both tools combined represent a total of \empirical{14.89} Ether, which is \emph{six orders of magnitude} less than the total. Second, for the classes of vulnerabilities treated by these tools, it is almost impossible to assess if these vulnerabilities have been exploited or not. Indeed, for a vulnerability such as being able to destruct the contract, there is no general way of knowing if the person who performed such a call should have been allowed to do so. Given this lack of reliability and the very low amount of money at stake, we decided to keep this data out of further analysis.

\point{Overlapping vulnerabilities}
In this subsection, we first check how much overlap there is between contracts in our dataset: how many contracts have been analyzed by multiple tools and how many contracts were flagged vulnerable by multiple tools.
We note that most papers, except for~\cite{Luu2016a}, are written around the same period. We find that~\empirical{73,627} out of~\empirical{821,219} contracts have been analyzed by at least two of the tools but only~\empirical{13,751} by at least three tools.
In Figure~\ref{fig:all-overlap}, we show a histogram of how many different tools analyze a single contract. In Figure~\ref{fig:vulnerable-overlap}, we show the number of tools which flag a single contract as vulnerable to any of the analyzed vulnerability. The overlap for both the analyzed and the vulnerable contracts is relatively small. We assume one of the reasons is that some tools work on Solidity code~\cite{DBLP:conf/ndss/KalraGDS18} while other tools work on EVM bytecode~\cite{Tsankov2018,Luu2016a}, making the population of contracts available different among tools.

We also find a lot of contradiction in the analysis of the different tools.
We choose re-entrancy to illustrate this point, as it is supported by three of the tools we analyze. In Figure~\ref{fig:reentrancy-agreement}, we show the agreement between the three tools supporting re-entrancy detection. The \emph{Total} column shows the total number of contracts analyzed by both tools in the \emph{Tools} column and flagged by at least one of them as vulnerable to re-entrancy. Oyente and Securify agree on only~\empirical{23\%} of the contracts, while Zeus does not seem to agree with any of the other tools.
This reflects the difficulty of building static analysis tools targeted at the EVM. While we are not trying to evaluate the different tools' performance, this gives us yet another motivation to find out the impact of the reported vulnerabilities.

% \begin{figure}[tb]
% \begin{lstlisting}[basicstyle=\footnotesize\ttfamily,language=json]
% [
%     {"op": "EQ", "pc": 7, 
%     "depth": 1, "stack": ["2b", "a3"]},
%     {"op": "ISZERO", "pc": 8, "depth": 1,
%     "stack": ["00"]}
% ]
% \end{lstlisting}
% \caption{\label{fig:execution-trace}Sample execution trace information.}
% \end{figure}

\section{Methodology}
\label{sec:methodology}

In this section, we describe in details the different analyses we perform in order to check for exploits of the vulnerabilities described in Section~\ref{sec:background}.

To check for potential exploits, we perform bytecode-level transaction analysis, whereby we look at the code executed by the contract when carrying out a particular transaction. We use this type of analysis to detect the~\VulnTypes types of vulnerabilities presented in Section~\ref{sec:background}.

To perform our analyses, we first retrieve transaction data for all the contracts in our dataset. Next, to perform bytecode-level analysis, we extract the execution traces for the transactions potentially affecting contracts of interest. We use EVM's debug functionality, which gives us the ability to replay transactions while tracing executed instructions. To speed-up the data collection process, we patch the Go Ethereum client~\cite{go-ethereum}, opposed to relying on the Remote Procedure Call~(RPC) functionality provided by the default Ethereum client.
% We show a truncated sample of the extracted traces in Figure~\ref{fig:execution-trace} for illustration. 
% %
% The \lstinline{op} key is the current instruction, \lstinline{pc} is the program counter, \lstinline{depth} is the current level of call nesting, and finally, \lstinline{stack} contains the current state of the stack. We use single-byte values in the example, but the actual values are~32 bytes~(256 bits).

The extracted traces contain a list of executed instructions, as well as the state of the stack at each instruction.
To analyze the traces, we encode them into a Datalog representation; Datalog is a language implementing first-order logic with recursion~\cite{Immerman99descriptivecomplexity}, allowing us to concisely express properties about the execution traces.
We use the following domains to encode the information about the traces as Datalog facts, noting $V$ as the set of program variables and $A$ is the set of Ethereum addresses.
We show an overview of the facts that we collect and the relations we use to check for possible exploits in Figure~\ref{fig:datalog-setup}.
We highlight that our analyses run \emph{automatically} based on the facts that we extract and the rules that define various violations described in subsequent sections. 

\begin{figure}
  \centering
  \begin{subfigure}{\columnwidth}
\begin{lstlisting}[language=Solidity,basicstyle=\footnotesize\ttfamily]
if (!addr.send(100)) { throw; }
\end{lstlisting}
    \caption{Failure handling in Solidity.}
    \label{fig:handled-exception-solidity}
  \end{subfigure}
  \begin{subfigure}{\columnwidth}
\begin{lstlisting}[language=esm, basicstyle=\footnotesize\ttfamily]
; preparing call
(0x65) CALL
; call result pushed on the stack
(0x69) PUSH1 0x73
(0x71) JUMPI ; jump to 0x73 if call was successful
(0x72) REVERT
(0x73) JUMPDEST
\end{lstlisting}
    \caption{EVM instructions for failure handling.}
    \label{fig:handled-exception-instructions}
  \end{subfigure}
  \caption{Correctly handled failed \lstinline{send}.}
  \label{fig:handled-exception}
\end{figure}

% \point{Analysis correctness: effects of soundness and completeness}
% Soundness and completeness affect our results in the following ways: if the analysis is \emph{unsound}, we might be missing exploits and therefore \emph{underestimating} the amount of Ether exploited.
% On the other hand, if the analysis is \emph{incomplete}, we might be flagging as exploit benign transactions and \emph{overestimating} the amount exploited.
% One of our goals in this work is to provide an upper-bound of the total amount exploited.
% Therefore, if a choice needs to be made between these two properties, we choose soundness over completeness for our analysis.

% \begin{figure}[tb]
%   \setlength{\tabcolsep}{3pt}
% \begin{tabular}{lcc}
% \toprule
% &	\bf Sound	& \bf Complete \\
% \midrule
% Re-entrancy	                  & yes	& yes\\
% Unhandled exception	          & yes	& yes\\
% Locked ether	                & yes	& yes\\
% Transaction order dependency	& yes	& no\\
% Integer overflow	            & no\footnotemark	& no\\
% \bottomrule\\
% \end{tabular}
% \caption{Soundness and completeness properties for the analyses in this paper.}
% \end{figure}

% \footnotetext{Section~\ref{ssec:method-io} and Section~\ref{ssec:analysis-io} provide in-depth explanations of the cases where the analysis is unsound}

%  is a set of statement identifiers;

\begin{figure}[tbp]
\begin{subfigure}[t]{\columnwidth}
  \centering
 \setlength{\tabcolsep}{4pt}
 \footnotesize
\begin{tabular}{ll}
  \toprule
  \bf Fact &  \bf Description\\
  \midrule
  \dterm{is_output}{v_1\in V\dsep v_2\in V} & $v_1$ is an output of $v_2$\\
  \dterm{size}{v\in V \dsep n\in \mathbb{N}} & $v$ has $n$ bits\\
  \dterm{is_signed}{v\in V} & $v$ is signed\\
  \dterm{in_condition}{v\in V} & $v$ is used in a condition \\
  \dterm{call}{a_1\in A\dsep a_2\in A\dsep p\in \mathbb{N}} & $a_1$ calls $a_2$ with $p$ Ether\\
  \dterm{create}{a_1\in A\dsep a_2\in A\dsep p\in \mathbb{N}} & $a_1$ creates $a_2$ with $p$ Ether\\
  \dterm{expected_result}{v\in V\dsep r\in \mathbb{Z}} & $v$'s expected result is $r$\\
  \dterm{actual_result}{v\in V\dsep r\in \mathbb{Z}} & $v$'s actual result is $r$\\
  \multirow{2}{*}{\dterm{call_result}{v\in V\dsep n\in\mathbb{N}}} & $v$ is the result of a call\\
  & and has a value of $n$\\
  \multirow{2}{*}{\dterm{call_entry}{i\in \mathbb{N}\dsep a\in A}} & contract $a$ is called when\\
           & program counter is $i$\\
  \multirow{2}{*}{\dterm{call_exit}{i\in \mathbb{N}}} & program counter is $i$ when\\
           & exiting a call to a contract\\
  \multirow{2}{*}{\dterm{tx_sstore}{b\in \mathbb{N}, i\in \mathbb{N}, k\in \mathbb{N}}} & storage key $k$ is written in\\
  & transaction $i$ of block $b$\\
  \multirow{2}{*}{\dterm{tx_sload}{b\in \mathbb{N}, i\in \mathbb{N}, k\in \mathbb{N}}} & storage key $k$ is read in\\
  & transaction $i$ of block $b$\\
  \dterm{caller}{v\in V, a\in A} & $v$ is the caller with address $a$\\
  \dterm{load_data}{v\in V} & $v$ contains transaction call data\\
  \dterm{restricted_inst}{v\in V} & $v$ is used by a restricted instruction\\
  \dterm{selfdestruct}{v\in V} & $v$ is used in \op{SELFDESTRUCT}\\
  \bottomrule
\end{tabular}
\caption{Datalog facts.}
\label{fig:datalog-facts}
\end{subfigure}

\vspace{2mm}

\begin{subfigure}[t]{\columnwidth}
  \centering
  \setlength{\tabcolsep}{1pt}
  \footnotesize
\begin{tabular}{l}
  \toprule
  \bf Datalog rules\\
  \midrule
  \dterm{depends}{v_1\in V\dsep v_2\in V} \drulesep \dterm{is_output}{v_1\dsep v_2}\dend\\
  \dterm{depends}{v_1\dsep v_2} \drulesep \dterm{is_output}{v_1\dsep v_3}\dsep \dterm{depends}{v_3\dsep v_2}\dend\\
  \midrule
  \dterm{call_flow}{a_1\in A\dsep a_2\in A\dsep p\in \mathbb{Z}} \drulesep \dterm{call}{a_1\dsep a_2\dsep p}\dend\\
  \dterm{call_flow}{a_1\in A\dsep a_2\in A\dsep p\in \mathbb{Z}} \drulesep \dterm{create}{a_1\dsep a_2\dsep p}\dend\\
  \dterm{call_flow}{a_1\dsep a_2\dsep p} \drulesep \dterm{call}{a_1\dsep a_3\dsep p}\dsep
  \dterm{call_flow}{a_3\dsep a_2\dsep \_}\dend\\
  \midrule
  \dterm{inferred_size}{v\in V\dsep n\in \mathbb{N}}\drulesep \dterm{size}{v\dsep n}\dend\\
  \dterm{inferred_size}{v\dsep n}\drulesep \dterm{depends}{v\dsep v_2}\dsep \dterm{size}{v_2\dsep n}\dend\\
  \midrule
  \dterm{inferred_signed}{v\in V}\drulesep \dterm{is_signed}{v}\dend\\
  \dterm{inferred_signed}{v}\drulesep \dterm{depends}{v\dsep v_2}\dsep \dterm{is_signed}{v_2}\dend\\
  \midrule
  \dterm{condition_flow}{v\in V\dsep v\in V}\drulesep \dterm{in_condition}{v}\dend\\
  \dterm{condition_flow}{v_1\dsep v_2} \drulesep \dterm{depends}{v_1\dsep v_2}\dsep
  \dterm{in_condition}{v_2}.\\
  \midrule
  \dterm{depends_caller}{v\in V}\drulesep \dterm{caller}{v_2\dsep \_}\dsep \dterm{depends}{v, v_2}.\\
  \midrule
  \dterm{depends_data}{v\in V}\drulesep \dterm{load_data}{v_2\dsep \_}\dsep \dterm{depends}{v, v_2}.\\
  \midrule
  \dterm{caller_checked}{v \in V}\drulesep \dterm{caller}{v_2, \_}\dsep\\
  \hspace{11.7em}\dterm{condition_flow}{v_2, v_3}\dsep $v_3 < v$.\\
  \bottomrule
\end{tabular}
\caption{Datalog rule definitions.}
\label{fig:relations}
\end{subfigure}

\vspace{2mm}

\begin{subfigure}[t]{\columnwidth}
  \centering
  \footnotesize
  \setlength{\tabcolsep}{1pt}
  \begin{tabular}{ll}
    \toprule
    \bf Vulnerability & \bf Query \\
    \midrule
    \reentrancy & \dterm{call_flow}{a_1\dsep a_2\dsep p_1}\dsep \\
                      & \dterm{call_flow}{a_2\dsep a_1\dsep p_2}\dsep $a_1 \neq a_2$\\
    \midrule
    Unhandled Excep. & \dterm{call_result}{v\dsep 0}\dsep $\lnot$\dterm{condition_flow}{v, \_}\\
    \midrule
    Transaction Order & \dterm{tx_sstore}{b, t_1, i}\dsep \\
    Dependency & \dterm{tx_sload}{b, t_2, i}, $t_1 \neq t_2$\\
    \midrule
    \lockedether & \dterm{call_entry}{i_1, a}\dsep \dterm{call_exit}{i_2}, $i_1 + 1 = i_2$\\
    \midrule
    \integeroverflow & \dterm{actual_result}{v\dsep r_1}\dsep\\
                              & \dterm{expected_result}{v\dsep r_2}\dsep$r_1 \neq r_2$\\
    \midrule
    \unrestrictedaction & \dterm{restricted_inst}{v}\dsep \dterm{depends_data}{v}\dsep\\
                      & $\lnot$\dterm{depends_caller}{v}\dsep $\lnot$\dterm{caller_checked}{v}\\
    & $\lor$ \dterm{selfdestruct}{v}\dsep $\lnot$\dterm{caller_checked}{v}\\
    \bottomrule
  \end{tabular}
  \caption{Datalog queries for detecting different vulnerability classes.}
\label{fig:queries}
\end{subfigure}
\caption{Datalog setup.}
\label{fig:datalog-setup}
\end{figure}

\subsection{\reentrancy}
In the EVM, as transactions are executed independently, re-entrancy issues can only occur \emph{within} a single transaction. Therefore, for re-entrancy to be exploited, there must be a call to an external contract which invokes, directly or indirectly, a re-entrant callback to the calling contract. We therefore start by looking for \op{CALL} instructions in the execution traces, while keeping track of the contract currently being executed.

When \op{CALL} is executed, the address of the contract to be called as well as the value to be sent can be retrieved by inspecting the values on the stack~\cite{wood2014ethereum}. Using this information, we can record \dterm{call}{a_1,a_2,p} facts described in Figure~\ref{fig:datalog-facts}.
We note that a contract can also create a new contract using \op{CREATE} and execute a re-entrancy attack using it~\cite{Rodler2019}. We therefore treat this instruction in a similar way as \op{CALL}.
Using these, we then use the query shown in Figure~\ref{fig:queries} to retrieve potentially malicious re-entrant calls.

\correctness Our analysis for re-entrant calls is sound but not complete. As the EVM executes each contract in a single thread, a re-entrant call must come from a recursive call. For example, given $A$, $B$, $C$ and $D$ being functions, a re-entrant call could be generated with a call path such as $A\rightarrow B \rightarrow C\rightarrow A$. Our tool searches for all mutually-recursive calls; it supports an arbitrarily-long calls path by using a recursive Datalog rule, making the analysis sound. However, we have no way of assessing if a re-entrant call is malicious or not, which can lead to false positives.

\subsection{\unhandledexceptions}
When Solidity compiles contracts, methods to send Ether, such as \lstinline{send}, are compiled into the EVM \op{CALL} instructions. We show an example of such a call and its instructions counterpart in Figure~\ref{fig:handled-exception}. If the address passed to \op{CALL} is an address, the EVM executes the code of the contract, otherwise it executes the necessary instructions to transfer Ether to the address. When the EVM is done executing, it pushes either $1$ on the stack, if the \op{CALL} succeeded, or $0$ otherwise. 

To retrieve information about call results, we can therefore check for \op{CALL} instructions and use the value pushed on the stack after the call execution. The end of the call execution can be easily found by checking when the \lstinline{depth} of the trace turns back to the value it had when the \op{CALL} instruction was executed; we save this information as \dterm{call_result}{v\dsep n} facts.
An important edge case to consider are calls to pre-compiled contracts, which are typically called by the compiler and do not require their return value to be checked, as they are results of computation, where $0$ could be a valid value, but could result in false-positives.
As pre-compiled contracts have known addresses between 1 and 10, we choose to simply not record \lstinline{call_result} facts for such calls.

As shown in Figure~\ref{fig:handled-exception-instructions}, the EVM uses the \op{JUMPI} instruction to perform conditional jumps. At the time of writing, this is the only instruction available to execute conditional control flow. We therefore mark all the values used as a condition in \op{JUMPI} as \lstinline{in_condition}. We can then check for the unhandled exceptions by looking for call results, which never influence a condition using the query shown in Figure~\ref{fig:queries}.

\correctness The analysis we perform to check for unhandled exceptions is complete but not sound.
All failed calls in the execution of the program will be recorded, while we accumulate facts about the execution.
We then use a recursive Datalog rule to check if the call result is used directly or indirectly in a condition.
We could obtain false negatives if the call result is used in a condition but the condition is not enough to prevent an exploit.
However, given that the most prevalent pattern for this vulnerability is the result of \lstinline{send} not being used at all~\cite{Tsankov2018}, and when the result is used, it is typically done within a \lstinline{require} or \lstinline{assert} expression, we hypothesize that such false negatives should be very rare.

\subsection{\lockedether}
Although there are several reasons for funds locked in a contract, we focus on the case where the contract relies on an external contract which does not exist anymore, as this is the pattern which had the largest financial impact on Ethereum~\cite{Breidenbach}. Such a case can occur when a contract uses another contract as a library to perform some actions on its behalf. To use a contract in this way, the \op{DELEGATECALL} instruction is used instead of \op{CALL}, as the latter does not preserve call data, such as the sender or the value.

The next important part is the behavior of the EVM when trying to call a contract which does not exist anymore.
When a contract is destructed, it is not completely removed per-se, but its code is not accessible anymore to callers.
When a contract tries to call a contract which has been destructed, the call is a no-op rather than a failure, which means that the next instruction will be executed and the call will be marked as successful.
To find such patterns, we collect Datalog facts about all the values of the program counter before and after every \lstinline{DELEGATECALL} instruction. In particular, we first mark the program counter value at which the call is executed~---~\dterm{call_entry}{i_1\in \mathbb{N}\dsep a\in A}. Then, using the same approach as for unhandled exceptions, we skip the content of the call and mark the program counter value at which the call returns~---~\dterm{call_exit}{i_2\in \mathbb{N}}.

If the called contract does not exist anymore, $i_1 + 1 = i_2$ must hold. Therefore, we can use the Datalog query shown in Figure~\ref{fig:queries} to retrieve the destructed contracts address.

\correctness The approach we use to detect locked Ether is sound and complete for the class of locked funds vulnerability we focus on. All vulnerable contracts must have a \lstinline{DELEGATECALL} instruction. If the issue is present and the call contract has indeed been destructed, it will always result in a no-op call. Our analysis records all of these calls and systematically check for the program counter before and after the execution, making the analysis sound and complete.

\subsection{\transactionorder}
The first insight to check for exploitation of transaction ordering dependency is that at least two transactions to the same contract must be included in the same block for such an attack to be successful. Furthermore, as shown in~\cite{Luu2016a} or~\cite{Tsankov2018}, exploiting a transaction ordering dependency vulnerability requires manipulation of the contract's storage.

The EVM has only one instruction to read from the storage, \op{SLOAD}, and one instruction to write to the storage, \op{SSTORE}. In the EVM, the location of the storage to use for both of these instructions is passed as an argument, and referred to as the storage \emph{key}. This key is available on the stack at the time the instruction is called. We go through all the transactions of the contracts and each time we encounter one of these instructions, we record either~\dterm{tx_sload}{b\in \mathbb{N}, i\in \mathbb{N}, k\in \mathbb{N}} or \dterm{tx_sstore}{b\in \mathbb{N}, i\in \mathbb{N}, k\in \mathbb{N}} where in each case $b$ is the block number, $i$ is the index of the transaction in the block and $k$ is the storage key being accessed.

The essence of the rule to check for transaction order dependency issues is then to look for patterns where at least two transactions are included in the same block with one of the transactions writing a key in the storage and another transaction reading the same key. We show the actual rule in Figure~\ref{fig:queries}.

\correctness Our approach to check for transaction order dependencies is sound but not complete. With the definition we use, for a contract to have a transaction order dependency it must have two transactions in the same block, which affect the same key in the storage. We check for all such cases, and therefore no false-negatives can exist. However, finding if there is a transaction order dependency requires more knowledge about how the storage is used and our approach could therefore result in false positives.

\subsection{\integeroverflow}
\label{ssec:method-io}

% \begin{figure}[tb]
%   \centering
%   \setlength{\tabcolsep}{14pt}
%   \begin{tabular}{rl}
%     \toprule
%     \textbf{Instruction} & \textbf{Description}\\
%     \midrule
%     \op{SIGNEXTEND} & Increase the number of bits\\
%     \op{SLT} & Signed lower than\\
%     \op{SGT} & Signed greater than\\
%     \op{SDIV} & Signed division\\
%     \op{SMOD} & Signed modulo\\
%     \bottomrule
%   \end{tabular}
%   \caption{EVM instructions that operate on signed operands.}
%   \label{fig:signed-instructions}
% \end{figure}
%
The EVM is completely untyped and expresses everything in terms of~256-bits words. Therefore, types are handled entirely at the compilation level and there is no explicit information about the original types in any execution traces.

To check for integer overflow, we accumulate facts over two passes. In the first pass, we try to recover the sign and size of the different values on the stack. To do so, we use known invariants about the Solidity compilation process. First, any value which is the result of an instruction such as \op{SIGNEXTEND} or \op{SDIV} can be marked to be signed with \dterm{is_signed}{v}. Furthermore, \op{SIGNEXTEND} being the usual sign extension operation for two's complement, it is passed both the value to extend and the number of bits of the value. This allows to retrieve the size of the signed value. We assume any value not explicitly marked as signed to be unsigned. To retrieve the size of unsigned values, we use another behavior of the Solidity compiler. 

To work around the lack of type in the EVM, the Solidity compiler inserts an \op{AND} instruction to ``cast'' unsigned integers to their correct value. For example, to emulate an \lstinline{uint8}, the compiler inserts \lstinline{AND value 0xff}. In the case of a ``cast'', the second operand $m$ will always be of the form $m = 16^n - 1,~n\in \mathbb{N},~n = 2^p,~p \in [1, 6]$. We use this observation to mark values with the according type: \lstinline{uintN} where $N = n \times 4$. Variables size are stored as \dterm{size}{v\dsep n} facts.

During the second phase, we use the \dterm{inferred_signed}{v} and \dterm{inferred_size}{v\dsep n} rules shown in Figure~\ref{fig:relations} to retrieve information about the current variable. When no information about the size can be inferred, we over-approximate it to $256$ bits, the size of an EVM word. Using this information, we compute the expected value for all arithmetic instructions (e.g. \op{ADD}, \op{MUL}), as well as the actual result computed by the EVM and store them as Datalog facts. Finally, we use the query shown in Figure~\ref{fig:queries} to find instructions which overflow.

\correctness Our analysis for integer overflow is neither sound nor complete. The types are inferred by using properties of the compiler using a heuristic which should work for most of cases but can fail. For example, if a contract contains code which yields \lstinline{AND value 0xff} but value is an \lstinline{uint32}, our type inference algorithm would wrongly infer that this variable is an \lstinline{uint8}. Such error during type inference could cause both false positives and false negatives. However, this type of issue occurs only when the developer uses bit manipulation with a mask similar to what the Solidity compiler generate. We find that such a pattern is rare enough not to skew our data, and give an estimate the possible number of contracts which could follow such a pattern in Section~\ref{ssec:analysis-io}.

\subsection{\unrestrictedaction}
\label{ssec:method-ua}
Unrestricted actions is a broad class of vulnerability, as it can include the ability to set an owner without being allowed to, destruct a contract without permission or yet execute arbitrary code.
As one of our main goal is to check the exploitation of vulnerable contracts, we stay close to the definitions given by previous works~\cite{217464} and focus on unrestricted Ether transfer using \op{CALL}, unrestricted writes using and \op{SSTORE}, and code injection using \op{DELEGATECALL} or \op{CALLCODE}.

First, we need to remind ourselves that the caller, unlike for example the call data, cannot be forged.
Therefore, one of the main insight is that if an action is restricted depending on who is calling, there should be an execution trace before the restricted operation which conditionally jumps, depending on the caller.
This is enough for \op{SELFDESTRUCT} but not for other instructions as it would flag a line such as \lstinline{balances[msg.sender] = msg.value} to be vulnerable.
To model this, we track whether the message sender influences the storage key or the address to call.
Finally, for code injection, we check whether the passed data influences the address called by \op{DELEGATECALL} or \op{CALLCODE}.

\correctness Our analysis for unrestricted actions is neither sound nor complete.
We take a relatively simple approach of checking whether the message sender influences a condition or not before executing a sensitive instruction.
This can result in false negatives because the check could be performed inappropriately, for example not reverting the transaction when needed, making the analysis unsound.
Furthermore, there might be some use cases where it is acceptable to allow any sender to write to the storage, but our analysis would flag such as vulnerable, resulting in false positives.
We discuss the implications further in Section~\ref{ssec:analysis-ua}.

\begin{figure}[tb]
  \centering
  \small
  \setlength{\tabcolsep}{2pt}
  \begin{tabular}{llr}
    \toprule
    \bf Contract address & \bf Last & \bf Amount \\
                         & \bf transaction & \bf exploited \\
    \midrule
    \addr[\scriptsize]{0xd654bdd32fc99471455e86c2e7f7d7b6437e9179} & 2016-06-10 & 5,885\\
    \addr[\scriptsize]{0x675e2c143295b8683b5aed421329c4df85f91b33} & 2015-12-31 & 50.49\\
    \addr[\scriptsize]{0xcd3e727275bc2f511822dc9a26bd7b0bbf161784} & 2017-03-25 & 10.34\\
    \bottomrule
  \end{tabular}
  \vskip 1mm
  \caption{\vre: Top contracts victim of re-entrancy attack and ETH amounts exploited}
  \label{fig:reentrancy-vulnerable}
\end{figure}

\section{Analysis of Individual Vulnerabilities}
\label{sec:analysis}

As described in Section~\ref{sec:datasets}, the combined amount of Ether contained within \emph{all} the vulnerable contracts exceeds~\empirical{3 million} ETH, worth~\empirical{\ToUSD{3} million} USD. In this section, we present the results for each vulnerability one by one; our results have been obtained using the methodology described in Section~\ref{sec:methodology}; the goal is to show how much of this money is actually at risk.

\point{Methodology}
\begingroup
\hypersetup{hidelinks}
For each vulnerability, we perform our analysis in two steps.
First, we fetch the execution traces of all the transactions up to block~\block{10200000} affecting the contracts in our dataset, either directly or through internal transactions. We then run our tool to automatically find the total amount of Ether at risk and report this number.
This is the amount we use to later give a total upper bound across all vulnerabilities.
In the second step, we manually analyze the contracts at risk to obtain more insight about the exploits and find interesting patterns.
As analyzing all the contracts manually would be impractical, for each vulnerability we manually analyze the contracts with the highest amount of Ether at risk to understand better the reasons behind the vulnerabilities.
We then present interesting findings as short case studies.
\endgroup

\point{Runtime performance} Our analysis runs in linear time and memory with respect to the number of instructions executed by a given transaction. The number of instructions varies widely between transactions, from a few hundreds to a few hundred thousands, with an average of around \empirical{100,000}. Our tool takes on average less than \empirical{10}ms (stddev. \empirical{20}ms) per transaction with a maximum of less than \empirical{2} seconds for the largest transactions, which is below the timeout of \empirical{5} seconds which we set for a single transaction.

% report all the contracts with a potential amount of Ether at risk greater than 10 ETH. We choose 10 ETH as a threshold because we think it is low enough to not miss any important attack, while being high enough to filter out most of the contracts. It is therefore appropriate for manual inspection effort. It is important to note that the filter used in this step affects in no way the total amount we report, and is used purely for practical purposes.

\subsection{\vre: \reentrancy}
\label{ssec:analysis-re}
There are~\empirical{4,337} contracts flagged as vulnerable to re-entrancy by~\cite{Luu2016a,Tsankov2018,DBLP:conf/ndss/KalraGDS18}, with a total of~\empirical{457,073} transactions. After running the analysis described in Section~\ref{sec:methodology} on all the transactions, we found a total of~\empirical{116} contracts which contain re-entrant calls. To look for the monetary amount at risk, we compute the sum of the Ether sent between two contracts in transactions containing re-entrant calls. The total amount of Ether exploited using re-entrancy is of~\empirical{6,076 ETH}, which is considerable as it represents more than \empirical{\ToUSD{6000} USD}.

\point{Manual analysis}
We manually analyze the top contracts in terms of fund lost and present them in Figure~\ref{fig:reentrancy-vulnerable}. Interestingly, one of these~\empirical{three} potential exploits has a substantial amount of Ether at stake:~\empirical{5,881 ETH}, which corresponds to around~\empirical{\ToUSD{5900} USD}. This address has already been detected as vulnerable by some recent work focusing on re-entrancy~\cite{Rodler2019}. It appears that the contract, which is part of the Maker DAO~\cite{maker-dao} platform, was found vulnerable by the authors of the contract, who themselves performed an attack to confirm the risk~\cite{ds-eth-token}.

\point{Sanity checking}
\begingroup
\hypersetup{hidelinks}
We use two different contracts for sanity checking.
First, we look at TheDAO attack, which is the most famous instance of a re-entrancy attack. Our tool detects the following re-entrancy pattern: \href{https://etherscan.io/address/0xc0ee9db1a9e07ca63e4ff0d5fb6f86bf68d47b89}{the malicious account} calls \href{https://etherscan.io/address/0xbb9bc244d798123fde783fcc1c72d3bb8c189413}{TheDAO main account}, TheDAO main account calls into \href{https://etherscan.io/address/0xd2e16a20dd7b1ae54fb0312209784478d069c7b0}{the reward account} and the reward account sends the reward to the malicious account, allowing it to perform the re-entrant call into TheDAO main account.
\endgroup

As another sanity check, we look at a contract called SpankChain~\cite{spank-chain}, which is known to recently have been compromised by a re-entrancy attack. We confirm that our approach successfully marks this contract as having been the victim of a re-entrancy attack and correctly identifies the attacker contract.

Finally, we note that our tool finds all the re-entrancy patterns presented by Sereum~\cite{Rodler2019}, including delegated and create-based re-entrancy\footnote{https://github.com/uni-due-syssec/eth-reentrancy-attack-patterns}.

\begin{figure}[tb]
  \centering
  \small
  \setlength{\tabcolsep}{8pt}
  \begin{tabular}{lr}
    \toprule
    \bf Contract address & \bf Amount at risk\\
    \midrule
    \addr[\scriptsize]{0x7011f3edc7fa43c81440f9f43a6458174113b162} & 56.70\\
    \addr[\scriptsize]{0xb336a86e2feb1e87a328fcb7dd4d04de3df254d0} & 42.27\\
    \addr[\scriptsize]{0xdcabd383a7c497069d0804070e4ba70ab6ecdd51} & 33.44\\
    \addr[\scriptsize]{0xfd2487cc0e5dce97f08be1bc8ef1dce8d5988b4d} & 21.60\\
    \addr[\scriptsize]{0x9e15f66b34edc3262796ef5e4d27139c931223f0} & 10.50\\
    \bottomrule
  \end{tabular}
  \vskip 1mm
  \caption{\vue: Top contracts affected by unhandled exceptions and ETH amounts at risk}
  \label{fig:unhandled-exceptions}
\end{figure}

\subsection{\vue: \unhandledexceptions}
\label{ssec:analysis-ue}
There are~\empirical{11,427} contracts flagged vulnerable to unhandled exceptions by~\cite{Tsankov2018,Luu2016a,DBLP:conf/ndss/KalraGDS18} for a total of more than~\empirical{3.4 million} transactions, which is \emph{an order of magnitude} larger than what we found for re-entrancy issues.

We find a total of~\empirical{264} contracts where failed calls have not been checked for, which represents roughly~\empirical{2\%} of the flagged contracts. The next goal is to find an upper bound on the amount of Ether at risk because of these unhandled exceptions. We define the upper bound on the money at risk to be the minimum value of the balance of the contract at the time of the unhandled exception and the total of Ether which have failed to be sent. We then sum the upper bound of all issues found to obtain a total upper bound. This gives us a total of \empirical{271.89} Ether at risk for unhandled exceptions.

\point{Manual analysis}
We manually analyze the top contracts and summarize their addresses and the amount at risk in Figure~\ref{fig:unhandled-exceptions}. The Solidity code is available for \empirical{three} of these contracts. We confirm that in all cases the issue came from a misuse of a low-level Solidity function such as \lstinline{send}.

\begin{investigation}{0x7011f3edc7fa43c81440f9f43a6458174113b162}
The contract {\small\addr{0x7011f3edc7fa43c81440f9f43a6458174113b162}} has failed to send a total of \empirical{52.90} Ether and currently still holds a balance of \empirical{69.3} Ether at the time of writing. After investigation, we find that the contract is an abandoned pyramid scheme~\cite{ethereum-pyramid}. The contract has a total of \empirical{21} calls which failed, all trying to send \empirical{2.7} Ether, which appears to have been the reward of the pyramid scheme at this point in time. Unfortunately, the code of this contract was not available for further inspection but we conclude that there is a high chance that some of the users in the pyramid scheme did not correctly obtain their reward because of this issue.
\end{investigation}

\subsection{\vle: \lockedether}
\label{ssec:analysis-le}
There are~\empirical{7,285} contracts flagged vulnerable to locked Ether by~\cite{Tsankov2018},~\cite{Grech2018},~\cite{Nikolic2018a} and~\cite{DBLP:conf/ndss/KalraGDS18}. The contracts hold a total value of more than~\empirical{1.4 million} ETH, which is worth more than \empirical{\ToUSD{1} million USD}. We analyze the transactions of the contracts that could potentially be locked by conducting the analysis described in the previous section. Our tool shows than \emph{none} of the contracts are actually affected by the pattern we check for~---~i.e., dependency on a contract which had been destructed.
We note that our tool currently only covers dependency on a destructed contract as a reason for locked Ether and patterns such as unbounded mass operation are not yet covered.

\point{Parity wallet}
Contracts affected by the Parity wallet (\addr{0x863df6bfa4469f3ead0be8f9f2aae51c91a907b4}) bug~\cite{Breidenbach} were not flagged by the tools we analyzed, and are therefore not present in our dataset.
As this is one of the most famous cases of locked Ether, we test our tool on the contracts affected by this bug.
To find the contracts, we simply have to use the Datalog query for locked Ether in Figure~\ref{fig:queries} and insert the value of the Parity wallet address as argument $a$. Our results for contracts affected by the Parity bug indeed matches what others had found in the past~\cite{parity-wallet-freeze}, with the contract at address~\addr{0x3bfc20f0b9afcace800d73d2191166ff16540258} having as much as \empirical{306,276 ETH} locked.
% \point{Transaction pattern analysis}
% It is worth pointing out that some tools, such as MadMax~\cite{Grech2018},  check for other types of issues, which could  also lock Ether. To try to check for such issues ourselves, we search for contracts with high monetary value, which have been inactive for a notably long period of time to see whether Ether is indeed locked.

% We find a total of~\empirical{15} contracts, which follow this pawttern. We show the~\empirical{5} contracts with the highest balance in Figure~\ref{fig:dependency}. We manually inspect the top \empirical{three} contracts, which contain a substantial amount of Ether, as well as the contracts, which have never \emph{sent} any Ether. These top three contracts are all implementing multi-sig wallets, which are typically used to store Ether for long periods of time, thus explaining the inactivity. After further manual inspection, we concluded that none of the contracts had been exploited, nor were exploitable.

% The first contract, which never sent any Ether, at address \addr{0x5a5eff38da95b0d58b6c616f2699168b480953c9} has its code publicly available. After inspection, it seems to be a ``lifelog'' and the fact it is not sending Ether seems to be there by-design; in other words, the funds are not locked. Although we were not able to inspect the other contract because its code was not available, we did not find any vulnerability report for this address.
\begin{figure}[tb]
  \centering
  \small
  \setlength{\tabcolsep}{3pt}
  \begin{tabular}{llr}
    \toprule
    \bf Contract address & \bf First issue & \bf Balance \\
    \midrule
    \addr[\scriptsize]{0x3da71558a40f63b960196cc0679847ff50fad22b} & 2016-09-06 & 13,818\\
    \addr[\scriptsize]{0xd79b4c6791784184e2755b2fc1659eaab0f80456} & 2016-05-03 & 2,013\\
    \addr[\scriptsize]{0xf45717552f12ef7cb65e95476f217ea008167ae3} & 2016-03-15 & 1,064\\
    \bottomrule
  \end{tabular}
  \vskip 1mm
  \caption{\textsf{TOD}: Top contracts potentially victim of transaction ordering dependency attack.}
  \label{fig:tod-vulnerable}
\end{figure}

\subsection{\vto: \transactionorder}
There are~\empirical{1,881} contracts flagged vulnerable to transaction ordering dependency by~\cite{Luu2016a} and~\cite{DBLP:conf/ndss/KalraGDS18}. We run the analysis described in Section~\ref{sec:methodology} on their \empirical{3,002,304} transactions and obtain a total of~\empirical{54} contracts potentially affected by transaction-order dependency. To estimate the amount of Ether at risk, we sum up the total value of Ether sent, including by internal transactions, during all the flagged transactions, resulting in a total of~\empirical{297.2 ETH} at risk of transaction-order dependency.

\point{Manual analysis}
For each contract, we find the block where transaction order dependency could have happened with the highest balance and report top with their balance at the time of the issue in Figure~\ref{fig:tod-vulnerable}. We manually investigated the contracts listed, they all had their source code available. We confirmed that in all the contracts, it was possible for a user to read and write to the same storage location within a single block. We inspected further \addr{0x3da71558a40f63b960196cc0679847ff50fad22b}, a contract called \lstinline{WithDrawChildDAO} and found that the read was simply for users to check their balance, making the issue benign.

\begin{figure*}[tb]
  \centering
  \setlength{\tabcolsep}{3pt}
  \begin{tabular}{|lrrr||rr||rr|}
    \hline
    
    \multicolumn{4}{|c||}{\bf Vulnerable} & \multicolumn{2}{c||}{\bf Exploited contracts} &
    \multicolumn{2}{c|}{\bf Exploited Ether} \\ \hline
    \bf Vuln. & \bf Vulnerable & \bf Total Ether & \bf Transactions & \bf Contracts & \bf \% of contracts & \bf Exploited & \bf \% of Ether \bigstrut[t]\\
    & \bf contracts & \bf at stake & \bf analyzed & \bf exploited & \bf exploited & \bf Ether & \bf exploited\\

    \hline\hline
    \vre & 4,337  & 1,518,067 & 457,073    & 116 &  2.68\% & 6,076 & 0.40\% \bigstrut[t] \\
    \vue & 11,427 & 419,418   & 3,400,960  & 264 & 2.31\% & 271.9 & 0.068\% \\
    \vle & 7,285  & 1,416,086 & 10,660,066 &   0 & 0\% & 0 & 0\%\\
    \vto & 1,881  & 302,679   & 3,002,304  &  54 & 3.72\% & 297.2 & 0.091\%\\
    \vio & 2,492  & 602,980   & 1,295,913  &  62 & 2.49\% & 1,842 & 0.31\%\\
    \vua & 5,163  & 580,927   & 3,871,770  &  42 &   0.813\%  &   0    & 0\% \\
    \hline
    \bf Total & \VulnerableContracts & \EtherStake & \NumAnalyzedTransactions & \NumExploitedContracts & \PercentExploitedContracts & \ExploitedEther & \PercentExploitedEther \bigstrut \\
    \hline
  \end{tabular}
  \vskip 1mm
  \caption{Understanding the exploitation of potentially vulnerable contracts.}
  \label{fig:findings-summary}
\end{figure*}

\subsection{\vio: \integeroverflow}
\label{ssec:analysis-io}
There are~\empirical{2,472} contracts flagged vulnerable to integer overflow, which accounts for a total of more than~\empirical{1.2 million transactions}. We run the approach we described in Section~\ref{sec:methodology} to search for actual occurrences of integer overflows.
It is worth noting that for integer overflow analysis we rely on properties of the Solidity compiler. To ensure that the contracts we analyze were compiled using Solidity, we fetched all the available source codes for contracts flagged vulnerable to integer overflow from Etherscan~\cite{etherscan}. Out of~\empirical{2,492} contracts,~\empirical{945} had their source code available and all of them were written in Solidity.

\point{Effects of our formulation}
As mentioned in Section~\ref{ssec:method-io}, some types of bit manipulation in Solidity contracts which could result in our type inference heuristic failing. We use the source codes we collected here to verify up to what extent this could affect our analysis. We find that bit manipulation by itself is already fairly rare in Solidity, with only~\empirical{244} out of the~\empirical{2,492} contracts we collected using any sort of bit manipulation. Furthermore, most of the contracts using bit manipulation were using it to manipulate a variable as a bit array, and only ever retrieved a single bit at a time. Such a pattern does not affect our analysis. We found only~\empirical{33} contracts which used \lstinline{0xFF} or similar values, which could actually affect our analysis. This represents about \empirical{1.3\%} of the number of contracts for which the source code was available.

We find a total of~\empirical{62} contracts with transactions where an integer overflow might have occurred.
To find the amount of Ether at stake, we analyze all the transactions which resulted in integer overflows. We approximate the amount by summing the total amount of Ether transferred in and out during a transaction containing an overflow. We find that the total of Ether at stake is~\empirical{1,842 ETH}. This is most likely an over-approximation but we use this value as our upper-bound.

\point{Manual analysis}
We inspect some of the results we obtained a little further to get a better sense of what kind of cases lead to overflows.
We find that a very frequent cause of overflow is rather underflow of unsigned values.
We highlight one of such cases in the following investigation.

\begin{investigation}{0xdcabd383a7c497069d0804070e4ba70ab6ecdd51}
  \begingroup
  \hypersetup{hidelinks}
This contract was flagged positive to both unhandled exceptions and integer overflow by our tool.
After inspection, it seems that at block height~\block{1364860}, the owner tried to reduce the fees but the unsigned value of the fees overflowed and became a huge number. Because of this issue, the contract was then trying to send large amount of Ether. 
This resulted in failed calls which happened not to be checked, hence the flag for unhandled exceptions.
\endgroup
\end{investigation}

\subsection{\unrestrictedaction}
\label{ssec:analysis-ua}
There is a total of \empirical{5,163} contracts flagged by~\cite{Tsankov2018,Nikolic2018a,217464} as vulnerable to unrestricted actions for a total of \empirical{3,871,770} transactions. We use the approach described in Section~\ref{ssec:method-ua} and find a total of \empirical{42} contracts having suffered of unrestricted actions, which were all non-restricted self-destructs, but none of them held Ether at the time of the exploit.

\point{Effects of our formulation} As mentioned in Section~\ref{ssec:method-ua}, this analysis is not sound, which means we need to be cautious about false positives.
A contract could have a check on the message sender which is incorrect and be exploited but not be flagged as such.
While we hypothesize that it is an edge case, it cannot be completely excluded.
However, having an automation method for such a check requires knowing the intent of the programmer, for example through specifications, which is out-of-scope of this work.
We therefore decide to inspect the contracts in our dataset in more details to understand better the level of exploitation.

\point{Manual analysis}
The tool teEther flags \emph{exploitable} contracts, as opposed to simply \emph{vulnerable} contracts.
Therefore, expect these contracts to be more likely to have been exploited and focus on these for our manual analysis.
We fetch all the historical balances of teEther contracts and retrieve the maximum amount held at any point in time and find the total of these to equal~\empirical{4,921} Ether.
However, we find that~\empirical{4,867} Ether belonged to~\empirical{48} different contracts with the exact same bytecode, and all had the same transaction pattern, which we describe in the following investigation.

\begin{investigation}{0xac54413f686927054a56d35415ba49618634e105}
  All contracts with a high historical monetary value found by teEther share the same bytecode, creator and transaction pattern as this contract.
  The contracts are created by \addr{0x15f889d2469d1be0e0699632d8d448f2178a7afe}, receive Ether from Kraken, an exchange, and send the same amount to \addr{0xd1bf1706306c7b667c67ffb5c1f76cc7637685bd} a couple of blocks later.
  We could not find further information about these addresses.
  We decompile the contract to understand how the contracts were exploitable and find that during the few blocks they held money, exploiting the contract would have been as simple as sending a transaction with the address to which to transfer the funds as argument.
  This is a very dangerous situation but because the Ether was usually sent within a minute to another address, an attacker would have needed to be very proactive and use advance tooling to exploit the contract.
  This illustrates well how a contract can be \emph{exploitable} but have little chance of being \emph{exploited} in practice.
\end{investigation}

\point{Sanity checking} As a sanity check, in addition to our test suite, we use one of the most famous instance of an unrestricted action: the destructed Parity wallet library contract at address~\addr{0x863df6bfa4469f3ead0be8f9f2aae51c91a907b4}. We analyze the transactions and successfully find an unrestricted store instruction in transaction~\tx{0x05f71e1b2cb4f03e547739db15d080fd30c989eda04d37ce6264c5686e0722c9}, which was used to take control of the wallet.

\subsection{Summary}
We summarize all our findings, including the number of contracts originally flagged, the amount of Ether at stake, and the amount \emph{actually exploited} in Figure~\ref{fig:findings-summary}. The \emph{Contracts exploited} column indicates the number of contracts which are flagged exploited and \emph{\% Contracts exploited} is the percentage of this number with respect to the number of contracts flagged vulnerable. The \emph{Exploited Ether} column shows the maximum amount of Ether that could have been exploited and the next column shows the percentage this amount represents compared to the total amount at stake. The \emph{Total} row accounts for contracts flagged with more than one vulnerability only once.

Overall, we find that the \emph{number of contracts exploited} is non negligible, with about \empirical{2\%} to \empirical{4\%} of vulnerable contracts exploited for \empirical{4} of the \VulnTypesNum vulnerabilities covered in our study.
However, it is important to note that the percentage of Ether exploited is an order of magnitude lower, with at most \empirical{0.4\%} of the Ether at stake exploited for re-entrancy.
This indicates that exploited contracts are usually low-value.
We will expand on this argument further in Section~\ref{sec:discussion}.

\section{Limitations}
\label{sec:limitations} 
In this section, we present the different limitations of our system, and describe how we try to mitigate them.

\point{Soundness vs Completeness} As for most tools such as this one, we are faced with the trade-off of soundness against completeness. Whenever possible we choose soundness over completeness --- three out of six of our analyses are sound. When we cannot have a sound analysis, we are careful to only leave out cases which are unlikely to generate many false negatives. In other words, we try as much as possible to reduce the number of false negatives, even if this means increasing the number of false positives. Indeed, the main goal of our system is to provide us an upper-bound of the amount of potentially exploited Ether, which make false negatives undesirable. Furthermore, we manually check the high-value contracts flagged as exploited, false-positives will not have an important influence on the final results. As an example of this trade-off, for re-entrancy we flag any contract which was called using a re-entrant call and lost funds in the process. However, in some cases, it could be an expected behavior and the funds could have been transferred to an address belonging to the same entity.

\point{Dataset} We only run our tool on the contracts included in our dataset, which means that we might be missing some exploits which actually occurred. Indeed, we did not have any contract affected by the Parity wallet bug nor had we the contract affected by TheDAO hack in the dataset. However, one of the main goal of this paper is to quantify what fraction of vulnerabilities discovered by analysis tools is exploited in practice and our current approach allows us to do exactly this.

\point{Other types of attacks} Our tool and analysis does not cover every existing attack to smart contracts. There are, for example, attacks targeting ERC-20 tokens~\cite{8802438}, or yet some other types of DoS attacks, such as wallet griefing~\cite{Grech2018}.
Furthermore, some detected ``exploits'' could be the results of Honeypots~\cite{236240} but our tool does not cover such cases.
Although it would be interesting to also cover such cases, we had to make a decision about the scope of the tool. Therefore, we focus on the vulnerabilities which have been the most covered in the literature, which we hypothesise is representative of how common the vulnerabilities are.

\begin{figure}[tb]
  \centering
  \begin{subfigure}{.49\columnwidth}
    \centering
    \includegraphics[width=\textwidth]{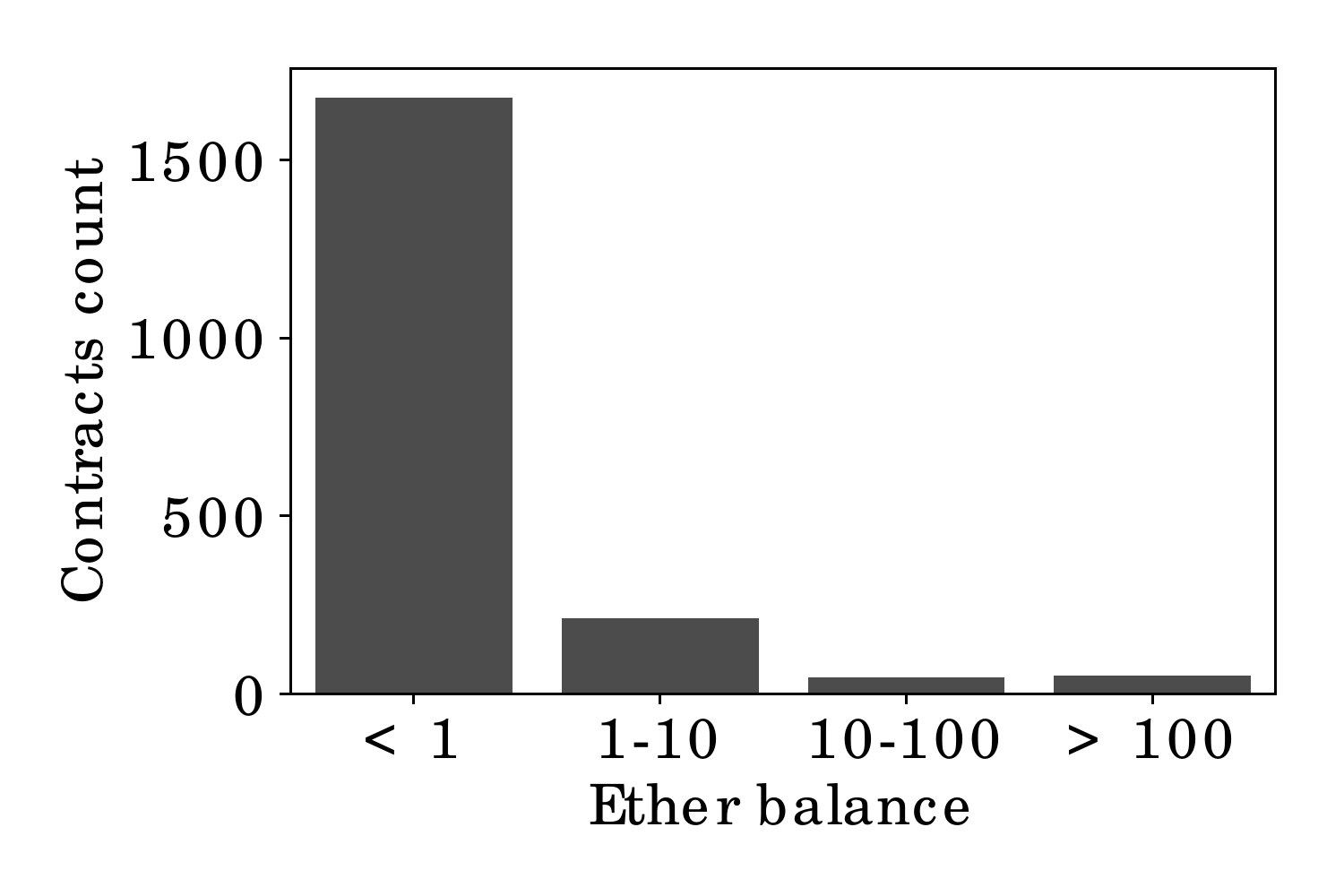}
    \caption{\scriptsize Ether held by contracts in our dataset\\with non-zero balance.}
    \label{fig:eth-held}
  \end{subfigure}
  \begin{subfigure}{.49\columnwidth}
    \centering
    \includegraphics[width=\textwidth]{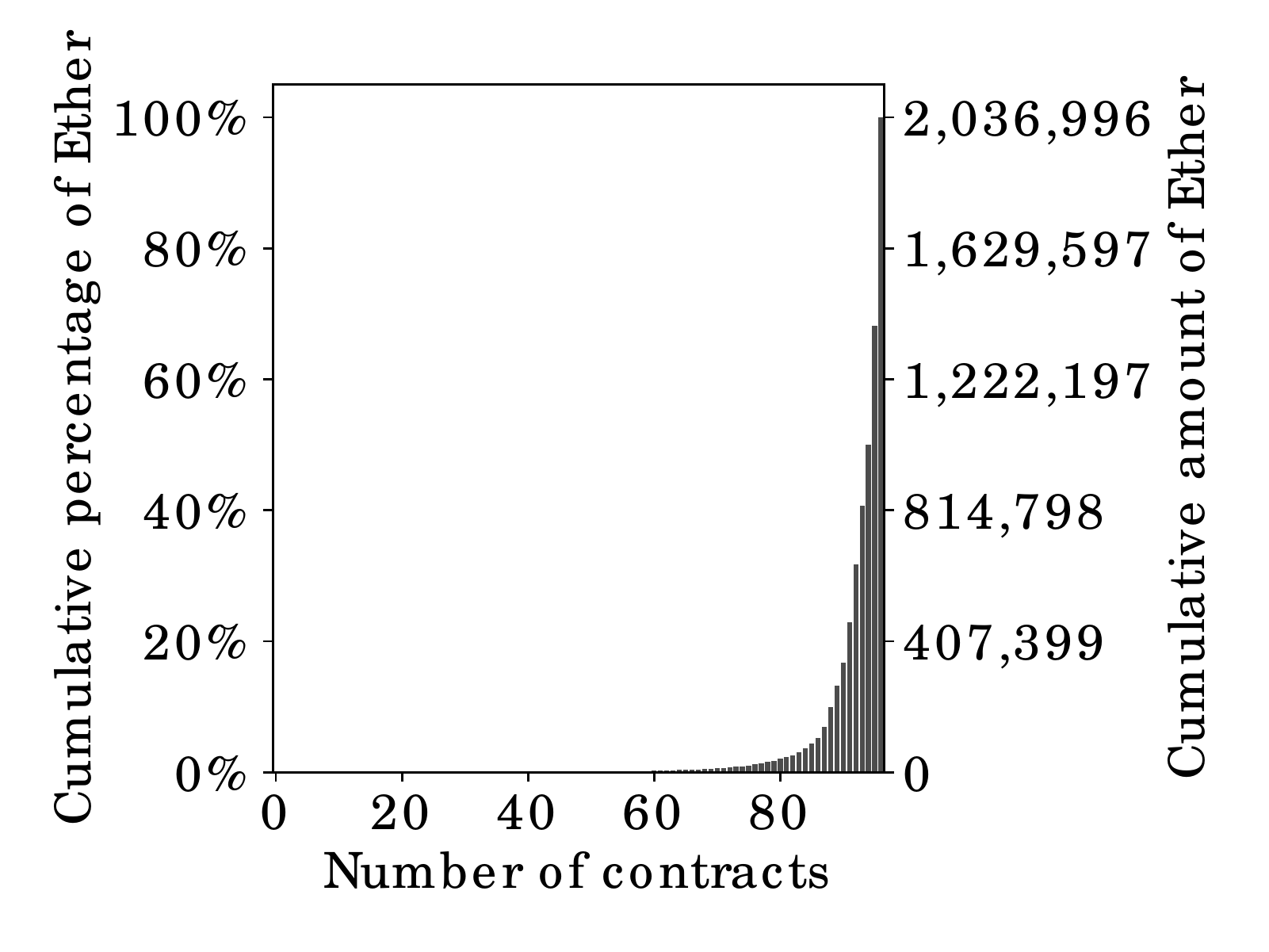}
    \caption{\scriptsize Cumulative Ether held in the 96\\contracts in our dataset containing at least 10 ETH.}
    \label{fig:cumulative-eth}
  \end{subfigure}
  \caption{Ether held in contracts: describing the distribution.}
\end{figure}

\section{Discussion}
\label{sec:discussion}

Even considering the limitations of our system, it is clear that the exploitation of smart contracts is vastly lower that what could be expected. In this section, we present some of the factors impacting the actual exploitation of smart contracts.

% \subsection{Contracts holding money}

We believe that a major reason for the difference between the number of vulnerable contracts reported and the number of contracts exploited is the distribution of Ether among contracts. Indeed, only about~\empirical{2,000} out of the~\VulnerableContracts contracts in our dataset contain Ether, and most of these contracts have a balance lower than~\empirical{1} ETH. We show the balance distribution of the contracts containing Ether in our dataset in Figure~\ref{fig:eth-held}. Furthermore, the \empirical{top 10 contracts} hold about~\empirical{95\%} of the total Ether. We show the cumulative distribution of Ether within the contracts containing more than~\empirical{10 ETH} in Figure~\ref{fig:cumulative-eth}. This shows that, as long as the top contracts cannot be exploited, the total amount of Ether that is actually at stake will be nowhere close to the upper bound of ``vulnerable'' Ether.

To make sure this fact generalizes to the whole Ethereum blockchain and not only our dataset, we fetch the balances of all existing contracts. This gives a total of~\empirical{15,459,193} contracts. Out of these, only~\empirical{463,538} contracts have a non-zero balance, which is merely 3\% of all the contracts. Out of the contracts with a non-zero balance, the top~10, top~100 and top~1000 account respectively for~\empirical{54\%},~\empirical{92\%} and ~\empirical{99\%} of the total amount of Ether. This shows that our dataset follows the same trend as the whole Ethereum blockchain: a very small amount of contracts hold most of the wealth.

\point{Manual inspection of high value contracts}
We decide to manually inspect the top~\empirical{6} contracts, in terms of balance at the time of writing, marked as vulnerable by any of the tools in our dataset. We focused on the top~\empirical{6} because it happened to be the number of contracts which currently hold more than~\empirical{100,000 ETH}. These contracts hold a total of~\empirical{1,695,240} ETH, or~\empirical{83\%} of the total of \empirical{2,037,521 ETH} currently held by all the contracts in our dataset. We give an overview of the findings here and a more in-depth version in Appendix~\ref{sec:investigations}.

\begin{figure*}[tb]
  \centering
  \small
  \setlength{\tabcolsep}{13pt}
  \begin{tabular}{lrll}
    \toprule
    \bf Address & \bf Ether balance & \bf Deployment date  & \bf Flagged vulnerabilities\\
    \midrule
    \addr[\footnotesize]{0xde0b295669a9fd93d5f28d9ec85e40f4cb697bae} & 649,493 & 2015-08-08 & Oyente: \vre\\
    \midrule
    \addr[\footnotesize]{0x7da82c7ab4771ff031b66538d2fb9b0b047f6cf9} & 369,023 & 2016-11-10 & MadMax:~\vle, Zeus:~\vio\\
    \midrule
    \addr[\footnotesize]{0x851b7f3ab81bd8df354f0d7640efcd7288553419} & 189,232 & 2017-04-18 & MadMax:~\vle\\
    \midrule
    \addr[\footnotesize]{0x07ee55aa48bb72dcc6e9d78256648910de513eca} & 182,524 & 2016-08-08 & Securify:~\vre\\
    \midrule
    \addr[\footnotesize]{0xcafe1a77e84698c83ca8931f54a755176ef75f2c} & 180,300 & 2017-06-04 & MadMax:~\vle\\
    \midrule
    \addr[\footnotesize]{0xbf4ed7b27f1d666546e30d74d50d173d20bca754} & 124,668 & 2016-07-16 & Securify:~\vto,~\vue; Zeus:~\vle,~\vio\\
    \bottomrule
  \end{tabular}
  \vskip 1mm
  \caption{Top \empirical{six} most valuable contracts flagged as vulnerable by at least one tool.}
  \label{fig:vulnerable-active}
\end{figure*}

\begin{investigation}{0xde0b295669a9fd93d5f28d9ec85e40f4cb697bae}
The source code for this contract is not available on Etherscan. However, we discovered that it is the multi-signature wallet of the Ethereum foundation~\cite{ether-foundation-contract-reddit} and that its source code is available on GitHub~\cite{ether-foundation-contract-code}. We inspect the code and find that all calls require the sender of the message to be an owner. This by itself is enough to prevent any re-entrant call, as the malicious contract would have to be an owner, which does not make sense. Furthermore, although the version of Oyente used in the paper reported the re-entrancy, more recent versions of the tool did not report this vulnerability anymore. Therefore, we safely conclude that the re-entrancy issue was a false alert.
\end{investigation}

We were able to inspect~\empirical{4} of the~\empirical{5} other contracts. The contract at address~\addr[\footnotesize]{0x07ee55aa48bb72dcc6e9d78256648910de513eca} is the only one for which we were unable to find any information. The second, third and fifth contracts in the list were also multi-signature wallets and exploitation would require a majority owner to be malicious. For example, for Ether to get locked, the owners would have to agree on adding enough extra owners so that all the loops over the owners result in an out-of-gas exception. The contract at address~\addr[\footnotesize]{0xbf4ed7b27f1d666546e30d74d50d173d20bca754} is a contract known as \lstinline{WithDrawDAO}~\cite{withdraw-dao}. We did not find any particular issue, but it does use a delegate pattern which explains the locked Ether vulnerability marked by Zeus.

We present a thorough investigation of the high-value contracts in Appendix~\ref{sec:investigations}. Overall, all the contracts from Figure~\ref{fig:vulnerable-active} that we could analyze seemed quite secure and the vulnerabilities flagged were definitely not exploitable. Although there are some very rare cases that we present in Section~\ref{sec:related} where contracts with high Ether balances are being stolen, these remain exceptions. The facts we presented up to now help us confirm that the amount of Ether at risk on the Ethereum blockchain is nowhere as close as what is claimed~\cite{DBLP:conf/ndss/KalraGDS18,Grech2018}.

% \subsection{Lifespan of contracts}
% As described in Section~\ref{sec:background}, due to smart contracts deployed on Ethereum being immutable, any update to a contract requires the deployment of a new one. As a result, smart contracts have a relatively short lifespan. We define the lifespan of a smart contract as the number of days elapsed between the first and the last transaction sent to it.

% \begin{figure}[tb]
%     % \includegraphics[width=\columnwidth]{./figures/recency.pdf}
%     \caption{Capturing the recency of active contract use.}
%     \label{fig:recency}
% \end{figure}
% Figure~\ref{fig:recency}...

\section{Related Work}
\label{sec:related}

Some major smart contracts exploits have been observed on Ethereum in recent years~\cite{Securities2017}. These attacks have been analyzed and classified~\cite{10.1007/978-3-662-54455-6_8} and many tools and techniques have emerged to prevent such attacks~\cite{harz2018towards,Dika2017}.
Recent literature has also shown how attacks on Ethereum are evolving with time~\cite{255252}.
In this section, we will first provide details about two of the most prominent historic exploits and then present existing work aimed at increasing smart contract security.

\subsection{Motivating Large-scale Exploits}
\point{TheDAO exploit}
TheDAO exploit~\cite{Securities2017} is one of the most infamous bugs on the Ethereum blockchain. Attackers exploited a re-entrancy vulnerability~\cite{10.1007/978-3-662-54455-6_8} of the contract which allowed for the draining of the contract's funds. The attacker contract could call the function to withdraw funds in a re-entrant manner before its balance on TheDAO was reduced, making it indeed possible to freely drain funds. A total of more than~3.5 million Ether were drained. Given the severity of the attack, the Ethereum community finally agreed on hard-forking.
\point{Parity wallet bug}
The Parity Wallet bug~\cite{Breidenbach} is another prominent vulnerability on the Ethereum blockchain which caused 280 million USD worth of Ethereum to be frozen on the Parity wallet account. It was due to a very simple vulnerability: a library contract used by the parity wallet was not initialized correctly and could be destructed by anyone. Once the library was destructed, any call to the Parity wallet would then fail, effectively locking all funds.

%\subsection{Other contract vulnerabilities}
%Recent research have proposed different taxonomies of smart contract vulnerabilities~\cite{Luu2016a,Atzei2017,Nikolic2018a}. Here, we will present a few classes of bug that enter in most of the proposed taxonomies.

% \subsection{Analysis Tools For Finding Vulnerabilities}

\subsection{Analyzing and Verifying Smart Contracts}
\label{ssec:related-analysis-tools}
There have been a lot of efforts in order to prevent such attacks and to make smart contracts more secure in general. We will here present some of the tools and techniques which have been presented in the literature and, when relevant, describe how they compare to our work.

Analysis tools can roughly be divided in two categories: static analysis and dynamic analysis tools.
Using the term ``static'' quite loosely, static analysis tools can be defined as tools which catch bugs or vulnerabilities without the need to deploy the smart contracts.
Runtime analysis tools try to detect these by executing the deployed contracts.
Our tool fits into the second category.

\point{Static analysis tools} Static analysis tools have been the main focus of research.
This is understandable, given how critical it is to avoid vulnerabilities in a deployed contract.
Most of these tools work by analyzing the bytecode or high-level code of contracts and checking for known vulnerable patterns.

Oyente~\cite{Luu2016a} is one of the first tools which has been developed to analyze smart contracts.
It uses symbolic execution in combination with the Z3 SMT solver~\cite{de2008z3} to check for the following vulnerabilities: transaction ordering dependency, re-entrancy and unhandled exceptions.

ZEUS~\cite{DBLP:conf/ndss/KalraGDS18} is a static analysis tool which works on the Solidity smart contract and not on the bytecode, making it appropriate to assist development efforts rather than to analyze deployed contracts, for which Solidity code is typically not available.
Zeus transpiles XACML-styled~\cite{XACML} policies to be enforced and the Solidity contract code into LLVM bitcode~\cite{lattner2004llvm} and uses constrained Horn clauses~\cite{bjorner2012program,mcmillan2007interpolants} over it to check that the policy is respected.

Securify~\cite{Tsankov2018} is a static analysis tool which checks security properties of the EVM bytecode of smart contracts.
It encodes security properties as patterns written in a Datalog-like~\cite{ullman1984principles} domain-specific language, and checks either for compliance or violation.
Securify infers semantic facts from the contract and interprets the security patterns to check for their violation or compliance by querying the inferred facts.
This approach has many similarities with ours, using Datalog to express vulnerability patterns.
The major difference is that Securify works on bytecode while our tool works on execution traces.

MadMax~\cite{Grech2018} has similarities with Securify, as it also encodes properties of the smart contract into Datalog, but it focuses on vulnerabilities related to gas.
It is the first tool to detect ``unbounded mass operations'', where a loop is bounded by a dynamic property such as the number of users, causing the contract to always run out of gas passed a certain number of users.
MadMax is built on top of the decompiler implemented by Vandal~\cite{DBLP:journals/corr/abs-1809-03981} and is performant enough to analyze all the contracts of the Ethereum blockchain in only~10 hours.

Several other static analysis tools have been developed, some, such as SmartCheck~\cite{Tikhomirov2017}, being quite generic and handling many classes of vulnerabilities, and other being more domain specific, such as Osiris~\cite{torres2018osiris} focusing on integer overflows, Maian~\cite{Nikolic2018a} on unrestricted actions or Gasper~\cite{Chen2017} on costly gas patterns.
More recently, ETHBMC~\cite{251546} was designed to also support inter-contract relations, cryptographic hash functions and memcopy-style operations.

Finally, there have also been some efforts to formally verify smart contracts. \cite{Hirai2017} is one of the first efforts in this direction and defines the EVM using Lem~\cite{mulligan2014lem}, which allows to generate definitions for theorem provers such as Coq~\cite{barras1997coq}. \cite{Grishchenko2018} presents a complete small-step semantics of EVM bytecode and formalizes it using the F* proof assistant~\cite{SwamyCFSBY11}. A similar effort is made in~\cite{Hildenbrandt2017} to give an executable formal specification of the EVM using the K Framework~\cite{rosu-serbanuta-2010-jlap}. VerX~\cite{permenev2019verx} is also a recent work allowing users to write properties about smart contracts which will be formally verified by the tool.

\point{Dynamic analysis tools} Although dynamic analysis tools have been less studied than their static counterpart, some work has emerged in recent years.

One of the first work in this line is ContractFuzzer~\cite{Jiang2018}.
As its name indicates, it uses fuzzing to find vulnerabilities in smart contracts and is capable of detecting a wide range of vulnerabilities such as re-entrancy, locked Ether or unhandled exceptions.
The tool generates inputs to the contract and checks using an instrumented EVM whether some vulnerabilities have been triggered.
An important limitation of this fuzzing approach is that it requires the Application Binary Interface of the contract, which is typically not available for contracts deployed on the main Ethereum network.

Sereum~\cite{Rodler2019} focuses on detecting re-entrancy exploitation at runtime by integrating checks in a modified Go Ethereum client.
The tool analyzes runtime traces and uses taint analysis to ensure that no variable accessing the contract storage is used in a re-entrant call.
Although there are some similarities with our tool, also analyzing traces at runtime, Sereum focuses on re-entrancy while our tool is more generic, notably because vulnerabilities pattern can easily be expressed using Datalog.

teEther~\cite{217464} also works at runtime but is different from the previous works presented, as it does not try to protect contracts but rather to actively find an exploit for them. It first analyzes the contract bytecode to look for critical execution paths.
Critical paths are execution paths which may result in lost funds, for example by sending money to an arbitrary address or being destructed by anyone.
To find these paths, it uses an approach close to Oyente~\cite{Luu2016a}, combining symbolic execution and Z3 to solve path constraints.

TXSPECTOR~\cite{255340}, which was published soon after the first version of this paper, uses a very similar approach to ours to detect re-entrancy, unchecked call and suicidal contracts.
They also leverage a Datalog approach to detect vulnerabilities but first transforms the transaction traces into a flow graph rather than adding facts about traces directly to the Datalog database.
While this does add expressiveness, it makes the analysis significantly more complex, resulting in some analysis timing out on some transactions. Therefore, we believe that their approach could be complementary to ours and used to eliminate potential false-positives of our approach.

\point{Summary} Static analysis tools are typically designed to detect \emph{vulnerable} contracts, while dynamic analysis tools are designed to detect \emph{exploitable} contracts. The only exception is Sereum, which detects contracts \emph{exploited} using re-entrancy.
Our work is, to the best of our knowledge, the first attempt to detect contracts \emph{exploited} using a wide range of vulnerabilities.
This is mostly orthogonal with other works and can support analysis tool development efforts by helping to understand what type of exploitation is happening in the wild.

\section{Conclusion}
\label{sec:conclusion}

In this paper, we surveyed the~\VulnerableContracts vulnerable contracts reported by~\PapersAnalyzed recent academic projects. We proposed a Datalog-based formulation for performing analysis over EVM execution traces and used it to analyze a total of more than~\empirical{20 million} transactions executed by these contracts. We found that at most~\NumExploitedContracts out of~\VulnerableContracts contracts have been subject to exploits but that at most~\ExploitedEther ETH (\ExploitedEtherUSD USD), or only~\PercentExploitedEther of the \EtherClaimedVulnerable ETH (\EtherClaimedVulnerableUSD USD) potentially at risk, was exploited.
Finally, we found that a majority of Ether is held by only a small number of contracts and that the vulnerabilities reported on these are either false positives or not exploitable in practice, thus providing a reasonable explanation for our results.

% Our results suggest that the impact of vulnerable smart contracts on the Ethereum blockchain had been exaggerated. We hypothesize that the main reasons for the significant gap between vulnerable and exploited are: many contracts flagged vulnerable are not in practice exploitable and most of the high-value contracts, which would be the most interesting to the attackers, fall into this category.

% \input{sections/acks}
%\newpage

% \bibliographystyle{ACM-Reference-Format}
\bibliographystyle{plain}
\bibliography{blockchain-security}
% \balancecolumns
\appendix
\section{Investigations}
\label{sec:investigations}
In this appendix, we will give a more in-depth security analysis of the top value contracts we presented in Section~\ref{sec:discussion}. In particular, we will focus on the vulnerabilities detected by the different tools and show how it could, or not, affect the contract.

\subsection*{\addr{0xde0b295669a9fd93d5f28d9ec85e40f4cb697bae}}
  This contract has been flagged as being vulnerable to re-entrancy by Oyente. For a contract to be victim of a re-entrancy attack, it must \lstinline{CALL} another contract, sending it enough gas to be able to perform the re-entrant call. In Solidity terms, this is means that the contract has to invoke \lstinline{address.call} and not explicitly set the gas limit. By looking at the source code~\cite{ether-foundation-contract-code}, we find 2 such instances: one at line 352 in the \lstinline{execute} function and another at line 369 in the \lstinline{confirm function}. The \lstinline{execute} is protected by the \lstinline{onlyowner} modifier, which requires the caller to be an owner of the wallet. This means that for a re-entrant call to work, the malicious contract would need to be an owner of the wallet in order to work. The \lstinline{confirm} function is protected by the \lstinline{onlymanyowners} modifier, which requires at least n owners to agree on confirming a particular transaction before it is executed, where n is agreed upon at contract creation time. Furthermore, \lstinline{confirm} will only invoke \lstinline{address.call} on a transaction previously created in the \lstinline{execute} function.

\subsection*{\addr{0x7da82c7ab4771ff031b66538d2fb9b0b047f6cf9}}
This is the contract for multi-signature wallet of the Golem project~\cite{golem-project} and uses a well-known multi-signature implementation. We use the source code available on Etherscan to perform the audit.
This contract is marked with locked Ether by MadMax and integer overflow by Zeus.

We first focus on the locked Ether which is due to an unbounded mass operation~\cite{Grech2018}. An unbounded mass operation is flagged when a loop is bounded by a variable which value could increase, for example the length of an array. This is because if the number of iteration becomes too large the contract would run out of gas every time, which could indeed result in locked funds. Therefore, we check all the loops in the contract. There are 8 loops in the code, at lines 43, 109, 184, 215, 234, 246, 257 and 265. All the loops except the ones at lines 257 and 265 are bound by the total number of owners. As owner can only be added if enough existing owners agree, running out-of-gas when looping on the number of owners cannot happen unless the owners agree. The loops at lines 257 and 265 are in a function called \lstinline{filterTransactions} and are bounded by the number of transactions. The function \lstinline{filterTransactions} is only used by two external getters, \lstinline{getPendingTransactions} and \lstinline{getExecutedTransactions} and could therefore not result the Ether getting lock. However, as the number of transactions is ever increasing, if the owner submit enough transactions, the \lstinline{filterTransactions} function could indeed need to loop over too many transactions and end up running out-of-gas on every execution. We estimate the amount of gas used in the loop to be around 50 gas, which means that if the number of transactions reaches 100,000, it would required more than 5,000,000 gas to list the transactions, which would probably make all calls run out of gas. The contract has only received a total of 281 transactions in more than 3 years so it is very unlikely that the number of transactions increase this much. Nevertheless, this is indeed an issue which should be fixed, most likely by limiting the maximum numbers of transactions that can be retrieved by \lstinline{getPendingTransactions} and \lstinline{getExecutedTransactions}.

Next, we look for possible integer overflows. All loops discussed above use an \lstinline{uint} as a loop index. In Solidity, \lstinline{uint} is a \lstinline{uint256} which makes it impossible to overflow here, given than neither the number of owners or transactions could ever reach such a number. The only other arithmetic operation performed is \lstinline{owners.length - 1} in the function \lstinline{removeOwner} at line 103. This function checks that the owner exists, which means that \lstinline{owners.length} will always be at least 1 and \lstinline{owners.length} can therefore never underflow.

\subsection*{\addr{0x851b7f3ab81bd8df354f0d7640efcd7288553419}}
This contract is also a multi-sig wallet, this time owned by Gnosis Ltd.\footnote{\url{https://gnosis.io/}} We use the source code available on Etherscan to perform the audit. The contract looks very similar of the one used by \addr{0x7da82c7ab4771ff031b66538d2fb9b0b047f6cf9} and has also been marked by MadMax as being vulnerable to locked Ether because of unbounded mass operations. Again, we look at all the loops in the contract and find that as the previous contract, it loops exclusively on owners and transactions. As in the previous contract, we assume looping on the owners is safe and look at the loops over the transactions. This contract has two functions looping over transactions, \lstinline{getTransactionCount} at line 303 and \lstinline{getTransactionIds} at line 351. Both functions are getters which are never called from within the contract. Therefore, no Ether could ever be locked because of this. Unlike the previous contract, \lstinline{getTransactionIds} allows to set the range of transactions to return, therefore making the function safe to unbounded mass operations. However, \lstinline{getTransactionCount} does loop over all the transactions, and as before, could therefore become unusable at some point, although it is highly unlikely.

\subsection*{\addr{0xcafe1a77e84698c83ca8931f54a755176ef75f2c}}
This contract is again a multi-sig wallet, this time owned by the Aragon project\footnote{\url{https://aragon.org/}}. We use the contract published on Etherscan for the audit. The source code for this contract is exactly the same as \addr{0x851b7f3ab81bd8df354f0d7640efcd7288553419}, except that it misses a contract called \lstinline[basicstyle=\ttfamily\small]{MultiSigWalletWithDailyLimit}. This contract was also flagged as being at risk of unbounded mass operations by MadMax, the conclusions are therefore exactly the same as for the previous contract.

\subsection*{\addr{0xbf4ed7b27f1d666546e30d74d50d173d20bca754}}
This contract is the only one which is very different from the previous ones. It is the \lstinline{WithdrawDAO} contract, which has been created for users to get their funds back after TheDAO incident~\cite{Securities2017}. We use the source code from Etherscan to audit the contract.
This contract has been flagged with several vulnerabilities: Securify flagged it with transaction order dependency and unhandled exception, while Zeus flagged it with locked ether and integer overflow.
The contract has two very short functions: \lstinline{withdraw} which allows users to convert their TheDAO tokens back to Ether, and the \lstinline{trusteeWithdraw} which allows to send funds which cannot be withdrawn by regular users to a trusted address.
We first look at the transaction order dependency. As any user will only ever be able to receive the amount of tokens he possesses, the order of the transaction should not be an issue in this contract. We then look at unhandled exceptions. There is indeed a call to \lstinline{send} in the \lstinline{trusteeWithdraw} which is not checked. Although it is not particularly an issue here, as this does not modify any other state, an error should probably be thrown if the call fails. We then look at locked ether. The contract is flagged with locked ether because of what Zeus classifies as ``failed send''. This issue was flagged because if the call to \lstinline{mainDAO.transferFrom} always raised, then the call to \lstinline{msg.sender.send} would never be reached, indeed preventing from reclaiming funds. However, in this context, \lstinline{mainDAO} is a trusted contract and it is therefore safe to assume that \lstinline{mainDAO.transferFrom} will not always fail. Finally, we look at the integer overflow issue. The only place where an overflow could occur is in \lstinline{trusteeWithdraw} at line 23. This could indeed overflow without some assumptions on the different values. For this particular contract, the following assumptions are made.
\vskip -7mm
\lstset{
  basicstyle=\ttfamily\footnotesize,
  mathescape
}
\begin{align*}
  &\text{\lstinline{this.balance}} \mbox{\footnotesize $+$} \text{\lstinline{mainDAO.balanceOf(this)}} \mbox{\footnotesize $\geq$} \text{\lstinline{mainDAO.totalSupply()}}\\
  &\text{\lstinline{mainDAO.totalSupply()}} \mbox{\footnotesize $>$} \text{\lstinline{mainDAO.balanceOf(this)}}
\end{align*}
\vskip -2mm
\noindent As long as these assumptions hold, which was the case when the contract was deployed, this expression will never overflow. Indeed, if we note $t$ the time before the first call to \lstinline{trusteeWithdraw} and $t + 1$ the time after the first call, we have
\begin{lstlisting}
this.balance$_{t+1}$ = this.balance$_t$ - (
  this.balance$_t$ + mainDAO.balanceOf(this)
                - mainDAO.totalSupply())
 = -mainDAO.balanceOf(this)+mainDAO.totalSupply()
\end{lstlisting}
\vskip -1mm
\noindent meaning that every subsequent call will compute:
\vskip -1mm
\begin{lstlisting}
this.balance$_{t+1}$ + mainDAO.balanceOf(this) -
                       mainDAO.totalSupply()
 = -mainDAO.balanceOf(this)+mainDAO.totalSupply()+
   mainDAO.balanceOf(this) - mainDAO.totalSupply()
 = 0
\end{lstlisting}
\vskip -3mm
\noindent This will always result in sending $0$ and will therefore not cause any overflow. If some money is newly received by the contract, the amount received will be transferred the next time \lstinline{trusteeWithdraw} is called.

% \input{sections/b_revision_reply}
% \balancecolumns
\end{document}